\title{Fast whole-slide cartography in colon cancer histology using superpixels and CNN classification}

\newcommand{\abstractText}{\noindent
\textbf{Purpose:} Automatic outlining of different tissue types in digitized histological specimen provides a basis for follow-up analyses and can potentially guide subsequent medical decisions. 
The immense size of \acp{wsi}, however, poses a challenge in terms of computation time. In this regard, the analysis of non-overlapping patches outperforms pixelwise segmentation approaches, but still leaves room for optimization. Furthermore, the division into patches, regardless of the biological structures they contain, is a drawback due to the loss of local dependencies. 
\\
\textbf{Approach:} We propose to subdivide the \ac{wsi} into coherent regions prior to classification by grouping visually similar adjacent pixels into superpixels. Afterwards, only a random subset of patches per superpixel is classified and patch labels are combined into a superpixel label. We propose a metric for identifying superpixels with an uncertain classification and evaluate two medical applications, namely tumor area and invasive margin estimation and tumor composition analysis.\\
\textbf{Results:} The algorithm has been developed on 159 hand-annotated \acp{wsi} of colon resections and its performance is compared to an analysis without prior segmentation. The algorithm shows an average speed-up of 41~\% and an increase in accuracy from 93.8~\% to 95.7~\%. By assigning a rejection label to uncertain superpixels, we further increase the accuracy by 0.4~\%. Whilst tumor area estimation shows high concordance to the annotated area, the analysis of tumor composition highlights limitations of our approach.\\
\textbf{Conclusion:} By combining superpixel segmentation and patch classification, we designed a fast and accurate framework for whole-slide cartography that is AI-model agnostic and provides the basis for various medical endpoints.
}

\documentclass[12pt, a4paper, twocolumn]{article}
\usepackage{xurl}
\usepackage[super,comma,sort&compress]{natbib}
\usepackage{abstract}

\usepackage{authblk}
\usepackage{graphicx}
\usepackage{amssymb}
\usepackage{amsmath}
\DeclareMathOperator*{\argmax}{arg\,max}
\usepackage{latexsym}
\usepackage{caption} 
\usepackage{float}
\usepackage{subfig}
\usepackage[small]{titlesec}
\usepackage[nolist]{acronym}

\usepackage{hyperref}
\hypersetup{colorlinks=true, urlcolor=blue, linkcolor=blue, citecolor=blue}

\author[$^{1,2}$]{Frauke Wilm}
\author[$^{\star,1}$]{Michaela Benz}
\author[$^1$]{Volker Bruns}
\author[$^1$]{Serop Baghdadlian}
\author[$^1$]{Jakob Dexl}
\author[$^1$]{David Hartmann}
\author[$^1$]{Petr Kuritcyn}
\author[$^1$]{Martin Weidenfeller}
\author[$^{1,2}$]{Thomas Wittenberg}
\author[$^3$]{Susanne Merkel}
\author[$^4$]{Arndt Hartmann}
\author[$^4$]{Markus Eckstein}
\author[$^4$]{Carol I. Geppert}

\affil[$^1$]{\small Fraunhofer Institute for Integrated Circuits IIS}
\affil[$^2$]{Friedrich-Alexander-University, Erlangen-Nuremberg}
\affil[$^3$]{Department of Surgery, University Hospital Erlangen, FAU Erlangen-Nuremberg}
\affil[$^4$]{Institute of Pathology, University Hospital Erlangen, FAU Erlangen-Nuremberg}
\affil[$\star$]{\textit{corresponding author: michaela.benz@iis.fraunhofer.de}}

\date{}

\begin{document}


\twocolumn[
  \begin{@twocolumnfalse}
    \maketitle
    \begin{abstract}
      \abstractText
      \newline
      \newline
    \end{abstract}
  \end{@twocolumnfalse}
]


\section{Introduction}
With the introduction of slide scanning systems into pathological workflows, the pre-requisite has been met to introduce machine learning algorithms into diagnostic routines. Due to their large size of over ten billion pixels, however, digitized histopathological \acfp{wsi} pose a challenge to automatic image analysis approaches. When working with such large images, technicians are oftentimes confronted with compromising computational efficiency for segmentation and classification accuracy. Especially in the clinical environment, however, both sides of the coin are equally desirable. This work focuses on how semantic segmentation of tissue classes can be executed efficiently. We present an algorithm for the analysis of large scale microscopic images which utilizes local pixel dependencies in order to achieve high classification accuracy, whilst maintaining reasonable computational complexity. We propose to introduce clustering into superpixels prior to classification which helps to model underlying biological structures. Furthermore, we present a technique of inferring superpixel classification labels using neural network classification. Using supervised learning and a hand-annotated database of 159 slides of colon resection specimens stained with \ac{he} dye, our solution is trained to distinguish seven tissue classes. The multi-class analysis of tissue facilitates a further evaluation of tumor composition and growth progression such as deriving the invasion front, which we only touch upon in this work, but do not cover in depth.

Beyond the general research question of how a whole-slide cartography can be performed efficiently, this work aims to answer the following more concrete questions. Can superpixel clustering prior to patch-based classification be utilized to achieve a speed-up? How large is the speed-up compared to a sole patch-based analysis and what is the impact on the segmentation accuracy? Does this approach work equally well for all tissue classes? Is it necessary and beneficial to classify all patches inside a superpixel or is it sufficient to classify only a subset? If so, what is the impact on the speed-up and accuracy and where is a good balance point? Considering medical end points, can the generated tissue map already be used to derive the tumor invasive margin? How accurately can the tumor area be calculated? Is the tumor composition (necrosis, active tumor cells, tumor stroma, mucus) accurately differentiated?

\section{Related work}
In the following, an overview of recent work in the field of semantic image segmentation and applications to pathological image data is provided. Furthermore, technically related approaches that combine superpixel clustering and subsequent classifications are briefly elaborated.

\subsection{Semantic segmentation}
Semantic image segmentation describes the process of inferring pixel-wise classification labels in order to generate a two-dimensional classification output. Due to their large size, \acp{wsi} are always divided into smaller image patches which are analyzed individually. Generally, two approaches for the semantic segmentation of \acp{wsi} can be distinguished: each image patch can be analyzed by a classification or segmentation network. The former predicts a single class-label for the whole image patch and after reassembling the classified patches, a segmentation mask of the \ac{wsi} can be obtained. This classification-based approach has been applied both in a non-overlapping manner \citep{cruz-roa2014, qaiser2016}, creating coarse segmentation masks, and, at the cost of higher computation times, in a sliding-window manner as neighborhood around each image pixel \citep{ciresan2012}. In order to incorporate image information on various scales, multiple resolutions can be integrated into a classification-based analysis \citep{wu2014, song2015, buyssens2012}. 

For the latter approach, based on the segmentation of image patches, special \ac{fcn} \citep{long2015} architectures such as U-Net \citep{oskal2019} or SegNet \citep{badrinarayanan2017} are typically used. These architectures employ encoder-decoder structures for the prediction of two-dimensional segmentation outputs and have been used for scene \citep{badrinarayanan2017} and biomedical image segmentation \citep{oskal2019, ronneberger2015,khened2021, mehta2018}. Encoder-decoder-based approaches are able to generate a segmentation output with a high granularity that can only be achieved by classification-based approaches when classifying each image pixel with its neighborhood as individual patches. However, these approaches entail high computational complexity and require extensive hardware resources.\citet{oskal2019}, for instance, reported inference times of up to 18 minutes per \ac{wsi} when using an NVIDIA Tesla P100 \acs{gpu} and \citet{khened2021} 30-75 minutes per \ac{wsi} with an NVIDIA Titan-V \acs{gpu}. These complex hardware requirements might not be attainable in a clinical setting and faster computation times are often desired.

\subsection{Applications in digital pathology}
In the field of digital pathology, machine learning algorithms have increasingly gained importance for answering pathological research questions. \citet{bychkov2018}, for instance, proposed a \ac{cnn}-based approach for directly predicting 5-year disease-specific survival for patients with colorectal cancer merely from tissue microarray cores. 

For the semantic segmentation of \acp{wsi}, two standard approaches can be distinguished: cell-based and texture-based methods.\citet{sirinukunwattana2016} designed a two-staged \ac{cnn}-based cell detection and classification algorithm, which has been utilized by various approaches \citep{sirinukunwattana2018, javed2018}. These incorporated graph structures to represent cell communities and thereby created phenotypic signatures. By splitting \acp{wsi} into smaller patches and mapping each to their most similar phenotypic signature, a multi-class \ac{wsi} cartography could be created. On colorectal cancer specimens, \citet{sirinukunwattana2018} scored an accuracy of 97.4~\% averaged over nine tissue classes and \citet{javed2018} an F$_1$~score of 92~\% averaged over six classes. These high classification scores, however, were achieved at the expense of high computation times of up to 50 minutes per \ac{wsi} for cell detection and classification \citep{sirinukunwattana2016}.

In the field of texture-based segmentation approaches \citet{signolle2008} proposed a method that incorporated several binary hidden wavelet domain Markov tree classifiers whose outputs were combined using majority voting. The authors scored a class-averaged recall of 71.02~\% on five tissue classes on ovarian carcinoma specimens with an inference time of up to 300 hours per \ac{wsi}. Other texture-based methods grouped pixels into coherent regions, which were classified using texture-based feature representations. On prostate specimens, \citet{gorelick2013} achieved a class-averaged recall of 83.88~\% on eight tissue classes with an inference time of two minutes per 300 x 300 pixel sized patch. \citet{apou2014} segmented breast cancer \ac{wsi} into six classes and achieved a class-averaged sensitivity of 55.83~\% and a class-averaged specificity of 91.4~\%. The authors stated inference times of under two hours per \ac{wsi}.

Due to the high variations in hardware resources and annotation quality, it is oftentimes difficult to compare image analysis algorithms in terms of classification accuracy and computational costs. \citet{kather2016} presented a publicly available dataset of histopathological image data and compared the performance of state-of-the-art image analysis algorithms. Using eight classes, the authors scored a maximum accuracy of 87.4~\% \citep{kather2016}. \citet{rachapudi2020} used this dataset to train a \ac{cnn} classifier and scored a class-averaged recall of 79.5~\% and a precision of 80.13~\%. Both, \citet{kather2016} and \citet{rachapudi2020}, however, achieved their quantitative results on test images that completely belonged to one class and the results are therefore difficult to compare to performance measures obtained on \acp{wsi} with multiple tissue classes present.

Using a binary cartography of histopathological images, primary tumor areas can be defined. Tumor here is defined as a combination of viable tumor cells, interconnecting tumor stroma and desmoplastic stroma as well as comprised necrotic areas and mucus. A robust definition of the tumor area can provide the basis for automatically evaluating pathological criteria such as tumor extend, composition or grading. Recent publications achieved tumor Dice scores of 69~\% \citep{balazsi2016} and 75.86~\% \citep{cruz-roa2017} on breast specimens and 78.2~\% \citep{khened2021} on colon samples. A good trade-off between refined segmentation results and low computational complexity was achieved by \citet{guo2019} who scored a tumor \ac{iou} value of 80.69~\% at an average inference time of 11.5~minutes per \ac{wsi}.

\subsection{Superpixel classification}
Due to their large size, digitized microscopic images can challenge standard machine learning algorithms. Aiming to reduce computational complexity, a clustering into coherent image segments, e.g. superpixels, has proven advantageous. \citet{zhang2021}, for instance, used superpixel clustering to compute a probability map for nuclei pre-segmentation, which was used as auxiliary input to the subsequent tissue classification network. \citet{nguyen2017} directly segmented breast tissue samples into coherent tissue regions using a graph-based superpixel algorithm. The authors, however, merely performed a segmentation and did not infer labels for the computed superpixels. Other existing works manually extracted hand-crafted superpixel feature vectors which were then classified using machine learning-based classifiers and thereby enabled a binary \citep{bejnordi2016,balazsi2016} or multi-class \citep{mehta2018,gorelick2013,apou2014,zormpas-petridis2018} semantic segmentation of medical images.  On histological image data, this approach has facilitated the binary segmentation of \acp{wsi} in 20-45~minutes by \citet{bejnordi2016} and up to 60~minutes by \citet{balazsi2016} with good performance results indicated by Dice scores of 92.43~\% \citep{bejnordi2016} and 69~\% \citep{balazsi2016}, respectively. \citet{mehta2018} segmented breast cancer tissues into eight classes by using superpixels and a \ac{svm} for classification. Since this combination was not the focus of their work, but merely served as a baseline for performance comparison of their proposed method, the usage of superpixels has not been evaluated in much detail. \citet{zormpas-petridis2018} applied a combination of superpixels and \ac{svm}-based classification on the task of segmenting melanoma \acp{wsi}. Their evaluation, however, was carried out with a randomly chosen set of superpixels, i.e. the ground truth did not contain the entire annotated tissues as in our work.

Considering the classification of image data, there has been a trend towards the use of deep learning methods, specifically \acp{cnn}, in recent years. \citet{bianconi2021} provided a comprehensive overview from theory-driven (hand-crafted) to data-driven (deep-learning) color and texture descriptors. \citet{Tamang2021} summarized various deep learning-based and classical approaches especially for the application of colorectal cancer diagnostics. One significant advantage of deep learning is that it enables a closed-form optimization of classification problems whereas classification based on hand-crafted features typically requires the selection of the most characteristic features followed by optimization of the classifier. In addition, \acp{cnn} often achieve more accurate classification results than traditional methods, especially when large amounts of labeled data are available for training, which was also shown in a comparison of different approaches for the classification of Malaria pathogens in microscopic image data made by \citet{Krappe2017}.

Due to their irregular size and shape, however, superpixels can challenge \ac{cnn} classifiers which require square input images of pre-defined size. Previous work in the field of histopathology can be categorized into two basic strategies to overcome this issue. The first group of approaches \citep{xu2016, turkki2016, zormpas2021, albayrak2018} extracted bounding boxes around superpixels and resized them to a pre-defined input size. This strategy either requires equally-sized superpixels to maintain a similar down-scaling factor for all superpixels or loses proportions across the input images. The latter can lead to ignoring the valuable size property of biological structures, e.g. the typically enhanced size of tumor cells, which can be an indicator for neoplastic growth. The second group of approaches \citep{sornapudi2018, pati2021} classified a pre-computed superpixel by extracting a patch with pre-defined size around the centroid of the superpixel. These approaches, however, relied on compact and square-like superpixels. Otherwise, the centroid might not lie within the given superpixel and the extracted patch will not be representative of this superpixel. Biological structures, however, are rarely square-shaped and especially at tumor boundaries the interaction of tumor, healthy tissue and inflammatory or necrotic reactions can lead to very irregularly shaped superpixels. In order to meet these characteristics of biological tissue and tumor growth, approaches that can be applied to superpixels of varying shapes and sizes are highly desired.  Moreover, all of these approaches \citep{xu2016, turkki2016, zormpas2021, albayrak2018,sornapudi2018, pati2021} relied on a one-to-one relationship between superpixel and the corresponding image patch, which is classified or processed by a \ac{cnn}. Only \citet{pati2021} subsequently merged neighboring and similar superpixels and averaged their \ac{cnn} feature vectors to use them in their tissue graph. In our approach, however, the superpixel shape is allowed to deviate greatly from a square shape, and the size of the superpixels is on average 20 times larger than the size of the image patches which are classified by the \ac{cnn}. This opens up the possibility of classifying multiple image patches within a superpixel and combining patch classification results to a superpixel label through majority voting. Moreover, this one-to-many relationship between superpixel and image patches allows deducing a classification confidence measure from the individual patch classification results.

\section{Material and methods}
The proposed image analysis pipeline has been trained and evaluated on colon \acp{wsi}, provided by the Institute of Pathology of the \ac{uker}. In the following sections an overview of the datasets and a detailed description of the applied methods is given.

\subsection{Datasets}
\begin{figure}[!ht]
    \centering
    \includegraphics[width=\columnwidth]{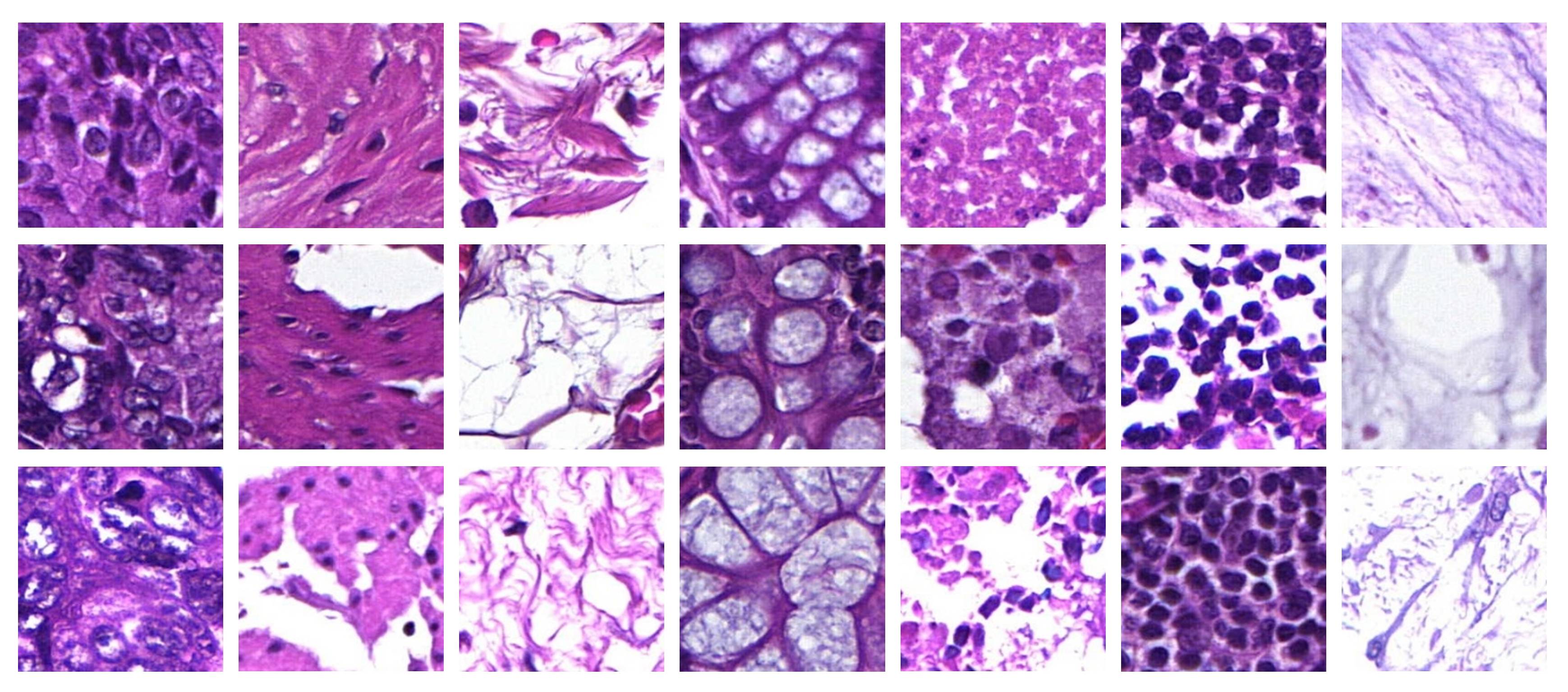}
    \caption{\label{fig:tissues}In each column representatives for one tissue class are displayed. From left to right:  tumor cells, muscle tissue, connective tissue combined with adipose tissue, mucosa, necrosis, inflammation, and mucus.}
\end{figure}

For this work, two different datasets have been used. Dataset A comprises 159 annotated \ac{he}-stained \acp{wsi}. The microscopic slides were digitized using a 3D HISTECH Pannoramic 250 slide scanner with an objective magnification of 20~X and a resolution of 0.22 x 0.22~$\mu$m / pixel. Pathologist-approved manual annotations cover seven tissue classes: tumor cells, muscle tissue, connective tissue combined with adipose tissue, mucosa, necrosis, inflammation, and mucus. Figure~\ref{fig:tissues} visualizes three representatives of each annotated class. Based on these annotations, patches of a size of 224 x 224 pixels that were covered to at least 85~\% by one annotation class have been extracted and labeled accordingly. These patches have been used for training and validating a neural network for semantic image segmentation. Table~\ref{tab:datasetA} provides an overview of the dataset including the total number of patches and the corresponding area. 

\begin{table}[!ht]
    \centering
    \caption{\label{tab:datasetA}{Overview of dataset A. The parameter test set is used to evaluate the superpixel configuration.}}
    \resizebox{\columnwidth}{!}{%
    \begin{tabular}{|lcrr|}
    \hline
        & \multicolumn{1}{c}{\# slides} & \multicolumn{1}{c}{\# patches} & \multicolumn{1}{c|}{area} \\ 
        & & \multicolumn{1}{c}{(224x224)} & \multicolumn{1}{c|}{($mm^2$)} \\
        \hline
        training set & 92 & 2,173,515 & 5,278\\
        validation set & 30 & 719,010 & 1,746\\
        parameter test set & 8 & - & 612\\
        test set & 29 & - & 3,047\\
        \hline
        sum & 159 & 2,892,525 & 10,683 \\
        \hline
    \end{tabular}}
\end{table}

A second dataset (dataset B) has been used for answering medical research questions, including tumor area estimation and composition. This dataset comprises 18 \ac{he}-stained samples with annotations of the primary tumor area and necrosis, inflammations and mucus within. Additionally, for each sample, an AE1/AE3 antibody \ac{ihc} as described by \citet{pourfarid2021} on a consecutive serial section was available. 

The retrospective study was approved by the scientific committee (CCC – tissue biobank) of the Comprehensive Cancer Center (CCC Erlangen-EMN; application-No. 100030; date of approval 09.05.2012) of the Friedrich-Alexander University Erlangen-Nuremberg. The study is based on the approval of the Ethics Commission of the University Hospital Erlangen (No. 4607 from 18.01.2012). The study is in accordance with the declaration of Helsinki and ethical guidelines applicable for retrospective studies were respected for all experiments. Tissue histology was reviewed by two pathologists. Pathology reports and medical records of patients who underwent an operation at our hospital were reviewed.

\subsection{Image analysis pipeline}
\label{sec:pipeline}
The developed image analysis pipeline is designed as a twofold approach: First, the \ac{wsi} is segmented into superpixels using the \ac{slic} algorithm \citep{achanta2012}. Then, each superpixel is classified using a \ac{cnn}-based approach.

\subsubsection{Superpixel segmentation} 
With the goal of reducing the computational complexity of a pixel-based clustering algorithm, the input \ac{wsi} is analyzed at a coarser resolution level (3.54~$\mu$m x 3.54~$\mu$m / pixel) corresponding to a down-scaling factor of 16 in each dimension with respect to the original resolution. Moreover, the \ac{wsi} is cropped at the tissue’s bounding box. The foreground (tissue) is determined by applying a simple intensity threshold to identify white background pixels (3.54~$\mu$m x 3.54~$\mu$m / pixel resolution). Afterwards, the remaining input image is segmented into superpixels. We compared different established superpixel clustering algorithms by \citet{achanta2012}, \citet{beucher1979}, \citet{felzenszwalb2004} and \citet{vedaldi2008}. These experiments demonstrated the superiority of the \ac{slic} algorithm regarding boundary detection of different tissue types and computational efficiency, which is in correspondence with the observations by \citet{achanta2012}. In this work we employ the \ac{slic} implementation from the Python scikit-image module. In order to utilize prior knowledge about the histological staining (\ac{he}), a color deconvolution \citep{ruifrok2001} is performed on the input image and the \ac{slic} algorithm has been modified by replacing the clustering in $[l,a,b,x,y]^T$-space with a clustering in $[H,E,x,y]^T$-space. In order to avoid overly jagged contours, the image is smoothed prior to segmentation using a Gaussian filter ($\sigma$=5). The \ac{slic}’s number of k-means iterations is limited to 10. The average superpixel size is set to 3,600 pixels at the down-scaled resolution level (i.e. a square superpixel would cover 0.2 x 0.2$~mm^2$). This average superpixel size was determined on a subset of Dataset A, which was solely used for parameter configuration (see Table~\ref{tab:datasetA}, chapter~\ref{sec:configurations}). Accordingly, the input parameter for the number of superpixels to be generated by the \ac{slic} algorithm is set to:

\begin{equation}
\resizebox{0.85\columnwidth}{!}{%
$k=\frac{pixelCount(boundingBox(foreground(WSI)))}{3,600}$
}
\label{equ:sizeS}
\end{equation}

\noindent After segmentation, all superpixels that contain at least 50~\% white pixels are labeled as background. These superpixels are excluded from any subsequent classification. The threshold of 50~\% has been set as a compromise in order to achieve an accurate tissue-background separation whilst not disregarding superpixels that cover adipose tissue, which oftentimes also contains large white areas.

\subsubsection{Superpixel classification} 
\begin{figure*}[!ht]
    \centering
    \includegraphics[width=0.8\textwidth]{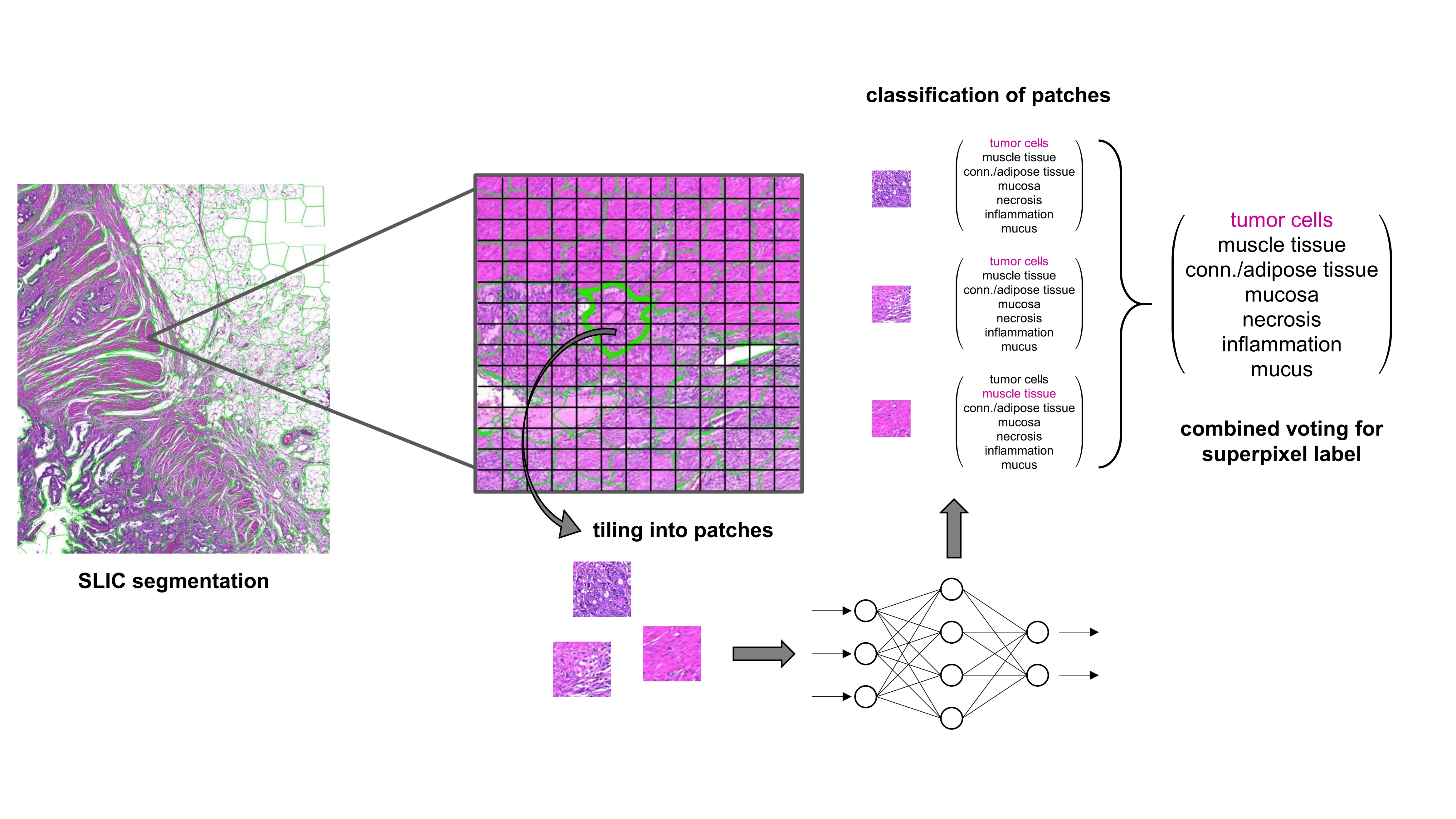}
    \caption{\label{fig:pipeline}Superpixel classification workflow}
\end{figure*}

Figure~\ref{fig:pipeline} visualizes the algorithm for inferring the superpixel class-labels. Initially, the input image is divided into equally sized, non-overlapping patches of 224 x 224 pixels at the original image resolution of 0.22 x 0.22~$\mu$m / pixel (20~X). Afterwards, all patches covered by at least 50~\% of one superpixel are classified using a \ac{cnn}. By lowering this threshold, the absolute number of patches that account to a superpixel’s classification result increases, but so does the relative number of in-distinctive border patches. The threshold of 50~\% was found to be a good trade-off during preliminary experiments on the parameter optimization subset of dataset A. After patch classification, all patch labels are combined to infer a superpixel classification. Various standard \ac{cnn} architectures utilize a softmax layer to output a probability distribution over all classes. We propose to compute the combined probability distribution by summing up over all patch softmax output vectors and normalizing by N, the number of patches that contribute to the classification result. The superpixel label $L_{SP}$ is then defined by the class corresponding to the maximum entry in the superpixel’s probability distribution:

\begin{equation}
\label{equ:LSP}
\resizebox{0.85\columnwidth}{!}{$%
\begin{split}
    L_{SP} & = \argmax_{c_i} \left( \frac{\sum_{n=1}^N {[p_{n,c_1},p_{n,c_2}, ...,p_{n,c_k}]^T}}{N} \right) \\
    & = \argmax_{c_i} \left( \frac{\sum_{n=1}^N \Vec{p_n}}{N} \right)
\end{split}$}%
\end{equation}

Here, $c_i \in C$ is the set of available class-labels.

Preliminary experiments have shown, that due to a high variance in shape and size of the superpixels sometimes up to 100 individual patches account to a superpixel label. Since the vast majority of patches within a superpixel will contain the same tissue type we hypothesize that valuable computation time can be saved by analyzing only a random subset of patches without significantly impacting the overall accuracy. We propose to analyze at most ten patches. The influence of this restriction is investigated in chapter~\ref{sec:configurations}

Moreover, we propose a confidence measure ($C_{diff}^{votes}$) of a superpixel classification derived from the classification results of the patches within this superpixel. For this, we divide the difference of patch votes for the most represented and the second most represented class by the number of all patches.

\subsubsection{CNN model}
The developed pre-processing steps (color deconvolution, foreground detection, superpixel segmentation) are independent of the subsequent \ac{cnn} structure which is therefore interchangeable. However, the average superpixel size has to be adapted to the \ac{cnn} input patch size in order to maintain a reasonable ratio of both measures. For the experiments elaborated hereafter a ResNet50 architecture with 224 x 224 pixel input size has been chosen and trained using training and validation set of Dataset A (see Table~\ref{tab:datasetA}). The network has been implemented using TensorFlow 2.2. We employ the color augmentation method described by \citet{tellez2018}, where the RGB image is converted to the \ac{he} color space using a deconvolution. Then, the Hematoxylin and Eosin components are individually modified, simulating different staining intensities. Moreover, zero-centering is applied as a preprocessing step. Training is performed using cross entropy loss and Adam optimizer with a learning rate of 0.001. A batch-size of 105 was chosen and in each batch the different classes are represented equally. Class imbalances are hereby compensated by oversampling of underrepresented classes as for example necrosis and mucus.

\subsection{Evaluation method for cartography results}
\label{sec:evaluation_methods_cartography}
For a visual validation of the annotation ground truths and classification outputs the Open Source software tool SlideRunner \citep{aubreville2018} has been used. The quantitative analysis is performed with an image resolution of 3.54~$\mu$m x 3.54~$\mu$m / pixel. We assign a class-label to each foreground pixel according to the manual ground truth annotation. The prediction map of the image is generated with the same resolution. Only pixels having both a ground truth and a prediction label are evaluated. Based on the confusion matrix different classification measures like e.g. class-wise recall are calculated. For all class-wise measures the corresponding two-class problem is regarded whereby all negative classes are combined to one class.

\subsection{Tumor area computation and invasive margin}
The primary tumor area is determined based on the cartography results. First, binary maps for the classes “tumor cells”, “necrosis” and “mucus” at the same resolution level used for superpixel segmentation (3.54~$\mu$m x 3.54~$\mu$m / pixel) are created. A morphological closing operation is applied to the “tumor cells” map and each connected component of the necrosis and mucus-map is checked for whether it is located adjacent to a tumor cell component. All adjacent necrosis and mucus components and all tumor cells components are added to a tumor map. Afterwards, morphological closing followed by opening is applied. Finally the tumor area is given by summing up areas enclosed by the outer contour of each tumor component. Besides the direct comparison of the calculated area (E) and the annotated ground truth area (GT), the \acf{iou} and Dice coefficient metrics are computed.

\begin{figure}[H]
    \centering
    \begin{equation}
    \label{equ:iou}
    \resizebox{0.85\columnwidth}{!}{%
        $IoU = \frac{|E \cap GT|}{|E \cup GT|} = \frac{TP}{TP+FP+FN}$} 
    \end{equation}
    \begin{equation}
    \label{equ:dice}
    \resizebox{0.85\columnwidth}{!}{%
        $Dice = \frac{2|E \cap GT|}{|E| + |GT|} = 
        \frac{2TP}{2TP + FP + FN}$}
    \end{equation}
    \caption*{Similarity measures (TP: true positives – pixel is contained in both segmented and annotated tumor area, FP: false positives – pixel is contained in segmented tumor area but not in the annotated one, FN: false negatives – annotated tumor pixel which is not contained in segmented tumor area)}
\end{figure}

The tumor invasive margin is derived from the auto-detected tumor area by extending the region in relation to the desired margin width. The intersection between this extended region and non-tumor tissue defines the basis of the invasive margin. Finally the intersection area is again extended as the invasive margin is situated at the border between tumor and surrounding tissue and stretching out into both.

\subsection{Tumor composition}
Quantitative analysis of the tumor micro-environment supports studies and diagnostics of \acp{til} in colon, as well as bladder and breast cancer \citep{dieci2018,pfannstiel2019}. The analysis in a region of interest like the invasive margin plays an important role for evaluating the immune response against tumor cells. In previous studies, a high correlation between CD3 and CD8 positive cell counts and patient outcome has been shown \citep{pages2018}. In the case of colon cancer, the immune response can be quantified using the immunoscore. 
 
\begin{figure}[!ht]
    \centering
    \subfloat{\includegraphics[width=.49\columnwidth]{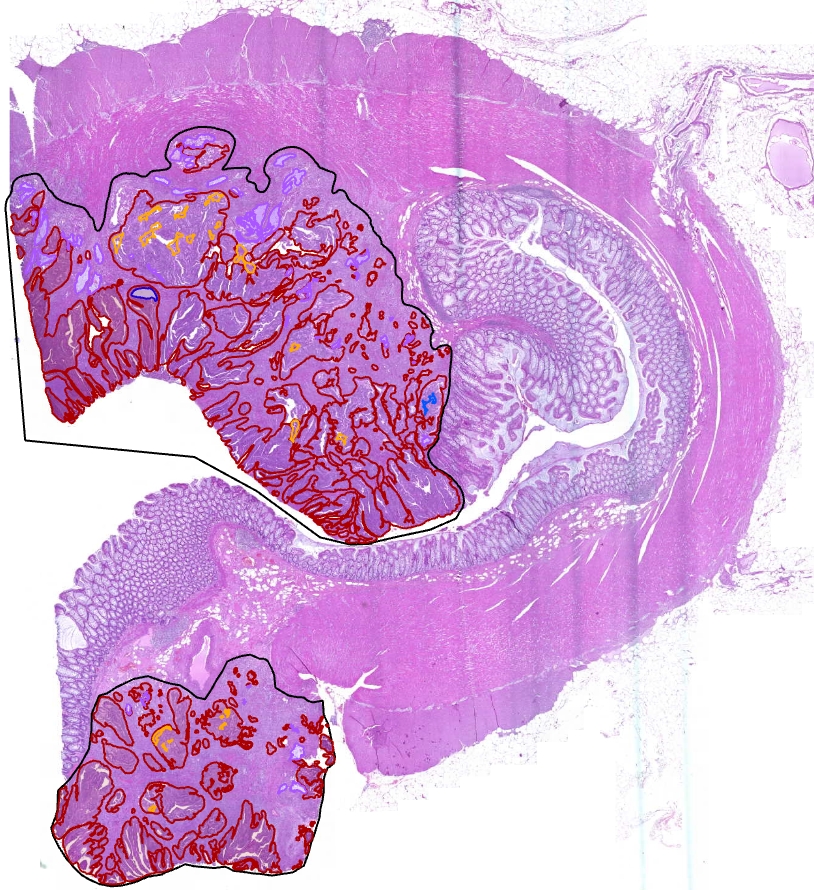}}
    \subfloat{\includegraphics[width=.49\columnwidth]{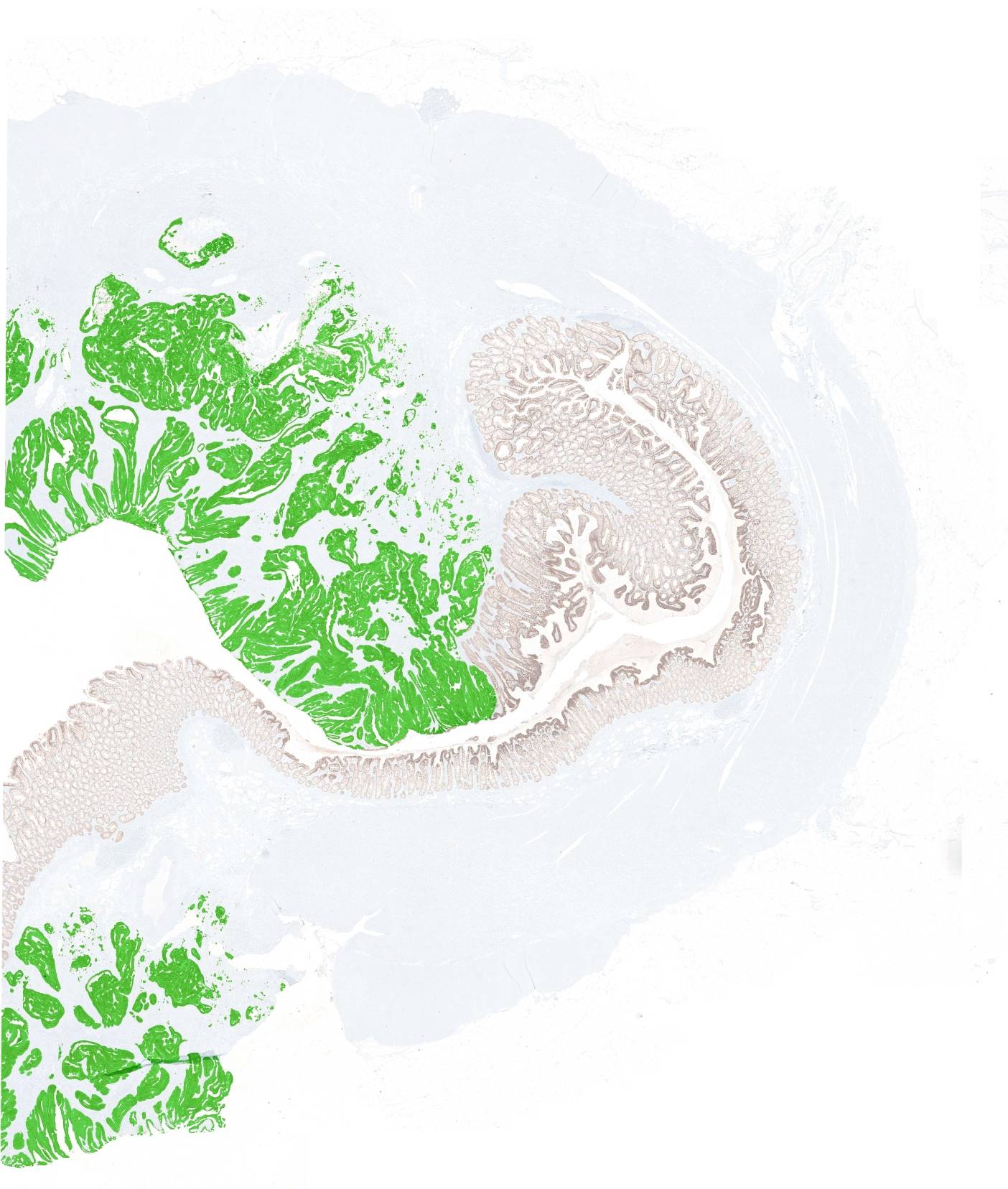}}
    \caption{\label{fig:HE_IHC}Manual annotation of tumor (black), necrosis (orange) and mucus (purple) in H\&E-stained colon section (left). Active tumor area (green) in corresponding IHC-stained colon section (right).}
\end{figure}

Therefore, we use dataset B to compare the estimation of tumor component areas (active tumor, necrosis, mucus) from cartography results with the manual annotations. \citet{graham2019} have used a rotation equivariant network for the task of gland segmentation. We use this approach to separate the active tumor area from interconnecting tumor stroma by using the segmented glands' area as approximation for the active tumor area. The ground truth area for necrosis and mucus is directly derived from the manual annotations. The ground truth for the active tumor area is obtained on serial sections stained with immunohistochemical markers (pan-cytokeratin, epithelial AE1/AE3) by applying a simple thresholding approach within the manually annotated tumor area. Again, color deconvolution was performed and solely the DAB channel was chosen for segmentation. Figure~\ref{fig:HE_IHC} shows a comparison of the manual annotations on the \ac{he}-stained \ac{wsi} and the segmentation result on the IHC-stained consecutive \ac{wsi}. On the one hand, this approach is beneficial as it does not suffer from the human annotator’s subjectivity. On the other hand, one has to keep in mind there is a small spatial distance between the two consecutive sections that is large enough that a cell visible in one slide might not be visible in the other slide.

\section{Results and discussion}
Several experiments were performed using dataset A to investigate the performance of the superpixel-based \ac{wsi} cartography. Starting from parameter configuration of the \ac{slic} algorithm, followed by a comparison between a classical patch-based approach with our newly introduced superpixel approach up to an investigation of uncertainty of the superpixel classification results. Though not the focus of this work, we carried out preliminary experiments on dataset B for two possible medical endpoints (tumor area and tumor composition) that will likely benefit from having a detailed tissue map available as it is generated by our proposed method.

\subsection{Configuration of superpixel approach}
\label{sec:configurations}
In order to define an optimal average superpixel size as well as a threshold for the number of classified patches per superpixel, experiments were performed on the parameter test set of dataset A containing eight \acp{wsi} (see Table~\ref{tab:datasetA}).

\begin{figure}[!ht]
    \centering
    \subfloat[Average number of superpixels per WSI and average computation time per superpixel classification.]{ \includegraphics[width=\columnwidth]{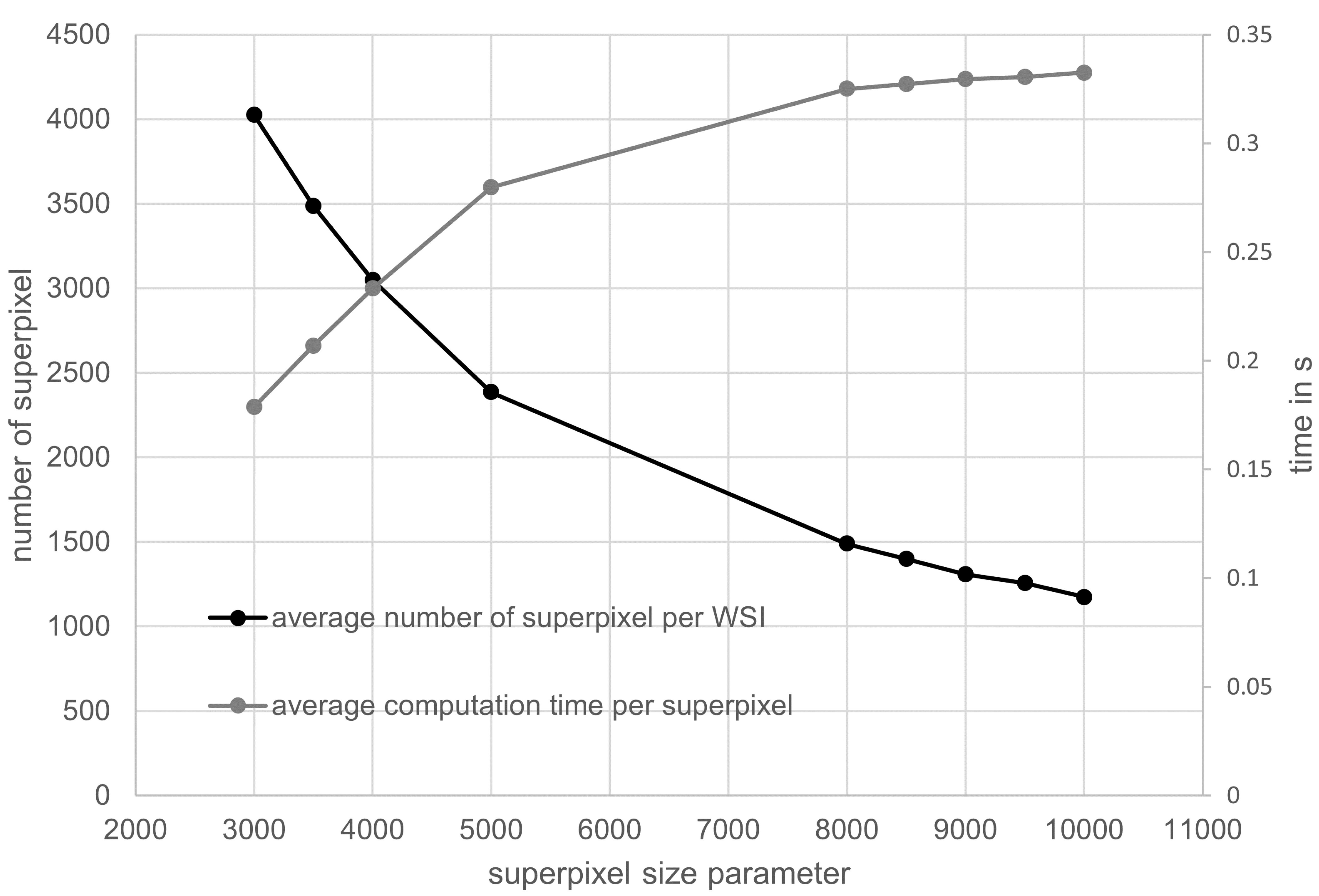}}
     \hfill
    \subfloat[Average computation time and accuracy for inference on overall slide.]{ \includegraphics[width=\columnwidth]{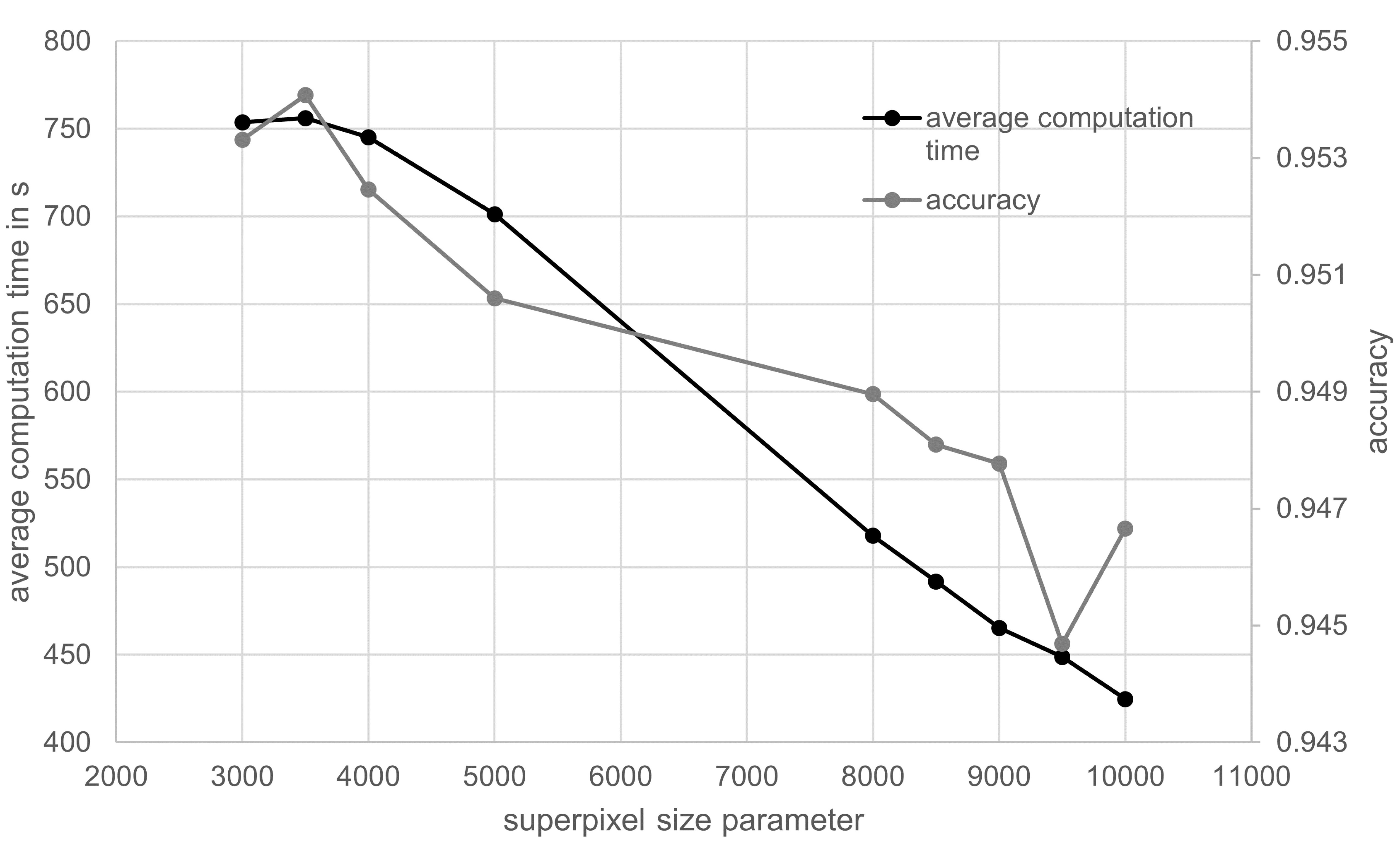}}
    \caption{\label{fig:influence_SP_size} Influence of average superpixel size on the average number of superpixels per \ac{wsi} and the average computation time per superpixel classification (left) and the average computation time and accuracy for inference on the overall slide (right). Evaluations were performed on the parameter test set of dataset A and the maximum number of classified patches per superpixel was limited to 30.}
\end{figure}

Figure~\ref{fig:influence_SP_size} visualizes the influence of increasing the average superpixel size (as start parameter for the \ac{slic} algorithm) on the total number of superpixels per \ac{wsi}, the classification accuracy and the average computation times for superpixel classification and \ac{wsi} inference. For these experiments, the maximum number of classified patches per superpixel was limited to 30. Inference times have been measured using an NVIDIA GeForce GTX 1060 \acs{gpu} with 6 GB RAM. As expected, a larger average size per superpixel results in fewer superpixels per \ac{wsi}. However, larger superpixels cover a larger number of patches, which are classified and then combined to infer a superpixel class-label. Therefore, larger superpixels entail higher computational costs (Figure~\ref{fig:influence_SP_size}a). Nevertheless, the overall computation time for slide inference decreases due to the decreased number of superpixels on the \ac{wsi}. The classification accuracy, however, also decreases (Figure~\ref{fig:influence_SP_size}b).

\begin{figure}[!ht]
    \centering
    \includegraphics[width=\columnwidth]{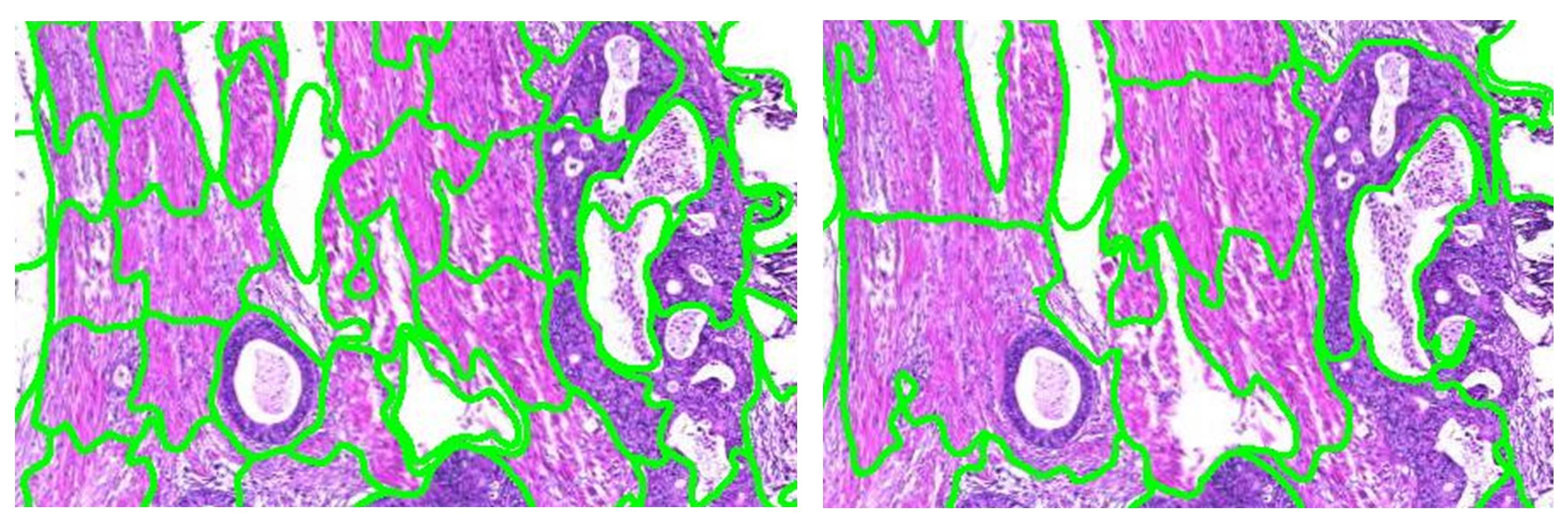}
    \caption{ \label{fig:superpixel_segmentation}Superpixel segmentation result with size of 3.600 pixels (left) compared to the segmentation result with a size of 10.000 pixels (right). The segmentation results on the left side fits better to the tumor cell boundaries but in both images necrotic areas within the tumor are not always detected as separate regions.}
\end{figure}

Figure~\ref{fig:superpixel_segmentation} visualizes the effect of smaller superpixel sizes (left) and larger superpixel sizes (right) on the segmentation result. As compromise between low computational complexity for larger superpixel sizes and high accuracy for smaller superpixel sizes, an average superpixel size of 3,600 pixels, i.e. a square superpixel would cover 0.2 x 0.2~mm$^2$, was chosen for further experiments. However, the results of these experiments depend on various parameters (such as the threshold for the number of classified patches per superpixel) as well as the chosen \ac{cnn} architecture and there is still room for further optimization. The biggest disadvantage of a greater superpixel size is that small details are neglected resulting in inaccurate segmentation results especially for classes like necrosis or tumor cells.

\begin{figure}[!ht]
    \centering
    \includegraphics[width=0.95\columnwidth]{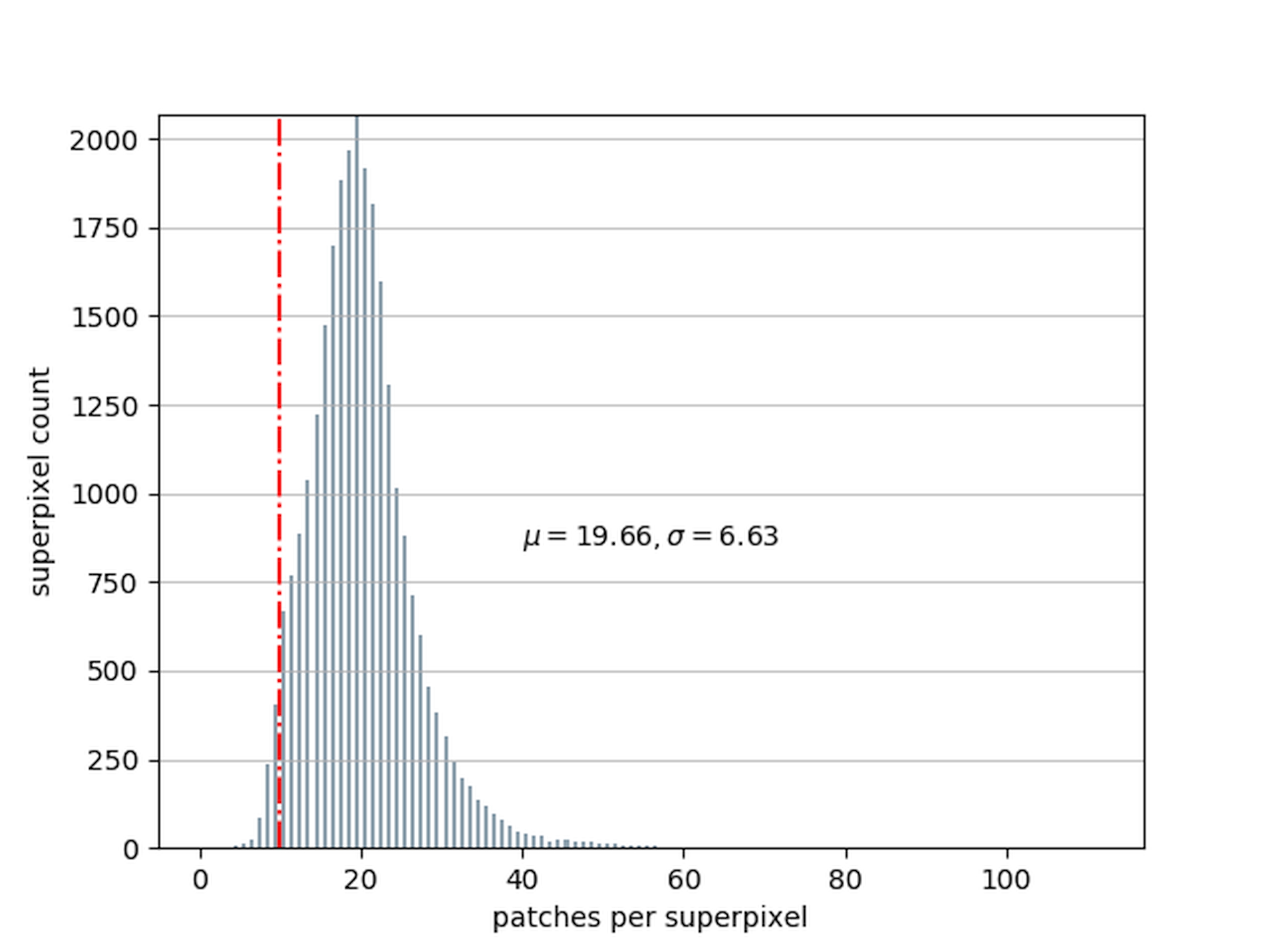}
    \caption{ \label{fig:histogram}Superpixel count over number of patches per superpixel for an average superpixel size of 3,600 pixels. The histogram sums up all \acp{wsi} in the parameter test set of dataset A. On average, a superpixel contains 19.66 patches with a standard deviation of 6.63.}
\end{figure}

The histogram in Figure~\ref{fig:histogram} shows, that with an average superpixel size of 3,600 pixels, some larger superpixels cover more than 30 individual patches. We hypothesized that it is sufficient to only classify a random subset of the patches within a superpixel. Table~\ref{tab:tile_limit} summarizes the influence of various maximum patch limits on the computation time of slide inference and overall accuracy. Whilst a smaller patch limit results in significantly lower computational costs, the slide accuracy only shows a marginal decrease. Therefore, we further reduced the limit from 30 to 10 patches for subsequent experiments.

\begin{table}[!ht]
    \caption{ \label{tab:tile_limit}Average computation time and overall accuracy on parameter test set for different limits of number of classified patches per superpixel.}
    \centering
\resizebox{\columnwidth}{!}{%
    \begin{tabular}{|ccc|}
        \hline
         patch restriction & average computation & overall \\
          per superpixel & time per WSI & accuracy \\
          \hline
         10 & 408 s & 95.1~\%\\
         18 & 643 s & 95.3~\%\\
         20 & 686 s & 95.3~\%\\
         \hline
    \end{tabular}}
\end{table}

\subsection{Classification performance and run-time}
In order to evaluate segmentation performance and computational complexity, the proposed algorithm is compared to a traditional classification-based approach with non-overlapping image patches. To isolate the effects produced by the proposed technique of introducing a superpixel clustering and inferring superpixel classification labels, the same \ac{cnn} is used as part of both approaches. Results are collected on the remaining 29 slides of dataset A (test set), which have not been used for training, validation or adaptation of parameters. The classification performance is assessed pixel-wise on a lower image resolution of 3.54~$\mu$m x 3.54~$\mu$m / pixel as described in chapter \ref{sec:evaluation_methods_cartography}. Table~\ref{tab:total_pixel} summarizes the total number of evaluated pixels on this resolution. Minor deviations of the overall sum of evaluated image pixels exist due to the irregular shape of the superpixels compared to the patch-wise approach.

\begin{table}[!ht]
    \centering
    \caption{\label{tab:total_pixel}Number of evaluated pixels (resolution 3.54~$\mu$m~x~3.54~$\mu$m~/~pixel) for the patch-based and superpixel approach. Differences are caused by the background detection and the irregular size of superpixels.}
    \resizebox{\columnwidth}{!}{%
    \begin{tabular}{|lrr|}
    \hline
    & \multicolumn{1}{c}{\# pixels} & \multicolumn{1}{c|}{\# pixels} \\
    & \multicolumn{1}{c}{(patch-based)}  &\multicolumn{1}{c|}{(superpixel)} \\
    \hline
    tumor cells & 61,575,137 & 61,526,076\\
    inflammation & 2,157,405 & 2,157,411\\
    conn./adipose tissue & 75,677,369 & 76,463,057\\
    muscle tissue & 60,759,062 & 60,652,301\\
    mucosa & 38,708,588 & 38,654,828\\
    mucus & 620,245 & 620,823\\
    necrosis & 5,883,073 & 5,874,393\\
    \hline
    sum & 245,380,879 & 245,948,889\\
    \hline
    \end{tabular}}
\end{table}

On the 29 test slides of dataset A, the tissue bounding box contains on average 10.7 billion pixels on the native resolution ($\hat{=}$ 520~mm$^2$). Within these, the \ac{slic} algorithm produces $4,060~\pm~1,717$ ($\mu\pm\sigma$) superpixels with an average size of 1,016,289 pixels ($\hat{=}$ 0.05~mm$^2$). The average number of patches per superpixel without introducing a maximum cut-off is $19.58~\pm~6.39$. A restriction of the maximum number of patches to be classified to only 10 patches per superpixel affects 94.8~\% of all superpixels and decreases the average number of classified patches per superpixel to $9.95~\pm~0.36$. 

\begin{figure}[!ht]
    \centering
    \subfloat[Superpixel-Based Approach] {\includegraphics[width=\columnwidth]{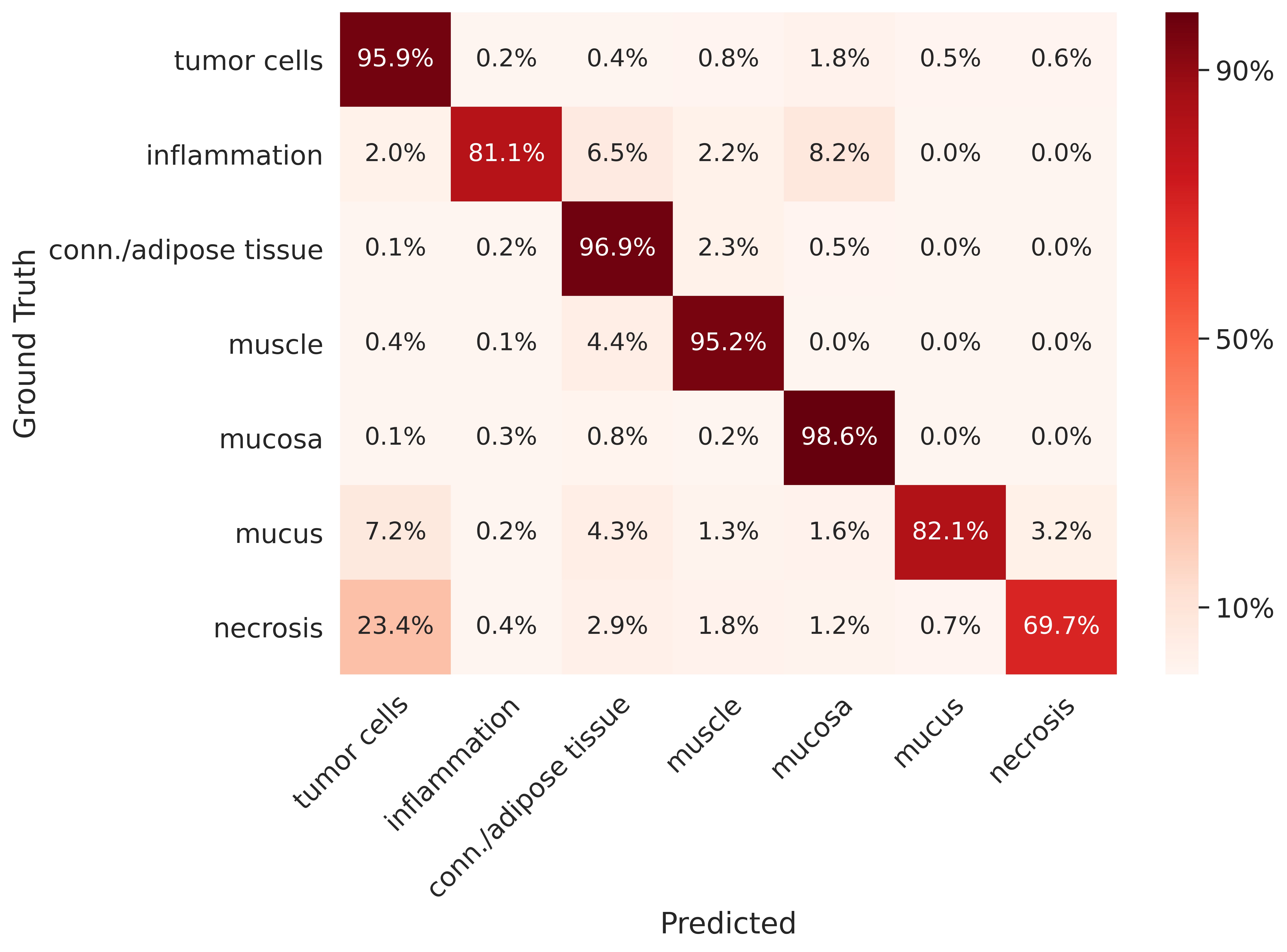}}
    \hfill
    \subfloat[Patch-Based Approach] {\includegraphics[width=\columnwidth]{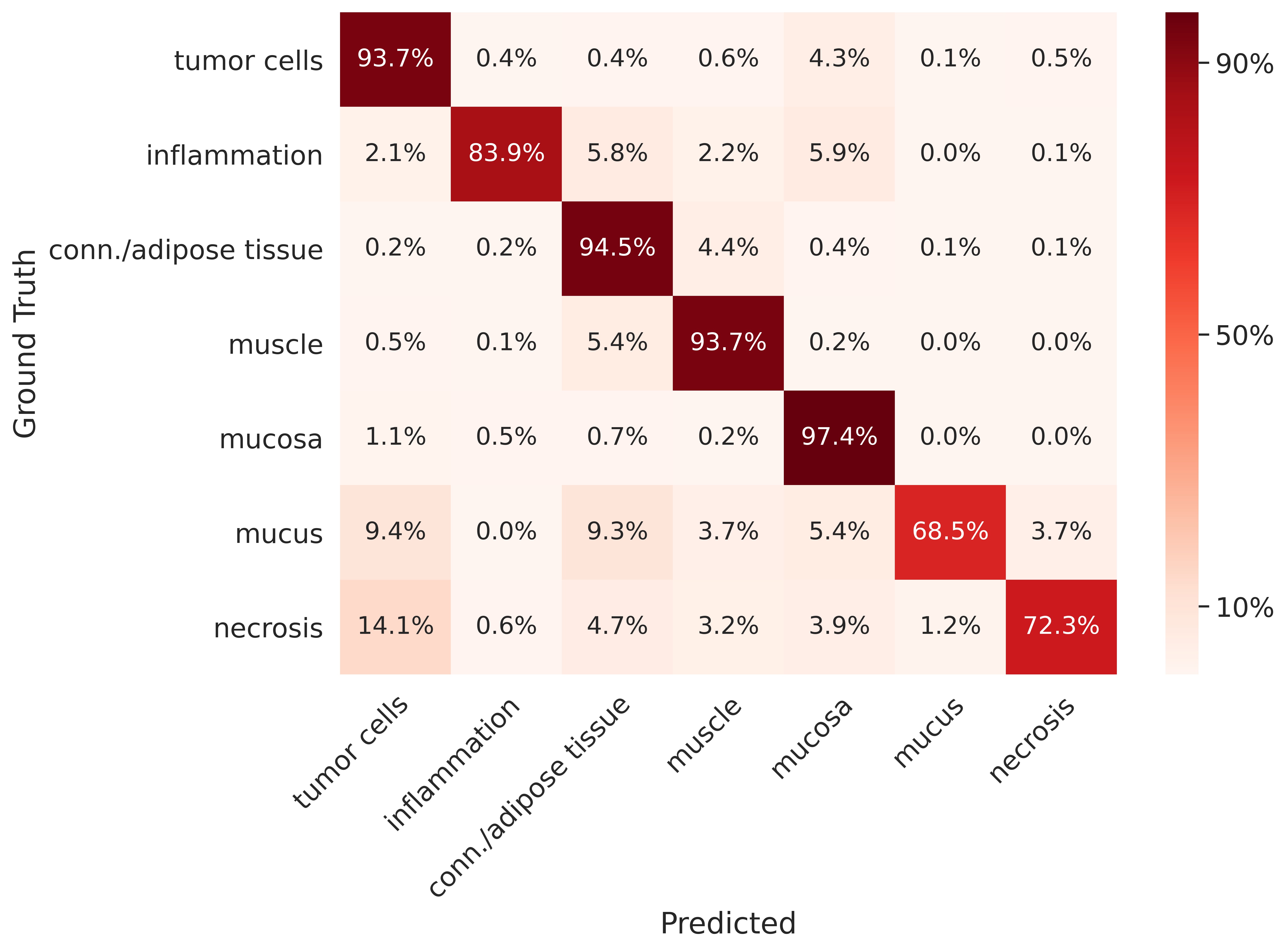}}
    \caption{\label{fig:confusion_matrices} Comparison of confusion matrices. The rows represent the ground truth class-labels and the columns represent the predictions. Due to high imbalances in the number of pixel per class a relative representation of the confusion matrix was chosen.}
\end{figure}

When evaluating a multi-class semantic segmentation task it is informative to look at which classes are frequently mistaken for one another. Figure~\ref{fig:confusion_matrices} shows the relative confusion matrices for both approaches. They show similar behavior regarding the typical confusions of classes: e.g. necrosis is misclassified as tumor or inflammation as mucosa. 

\begin{figure}[!ht]
    \centering
    \includegraphics[width=\columnwidth]{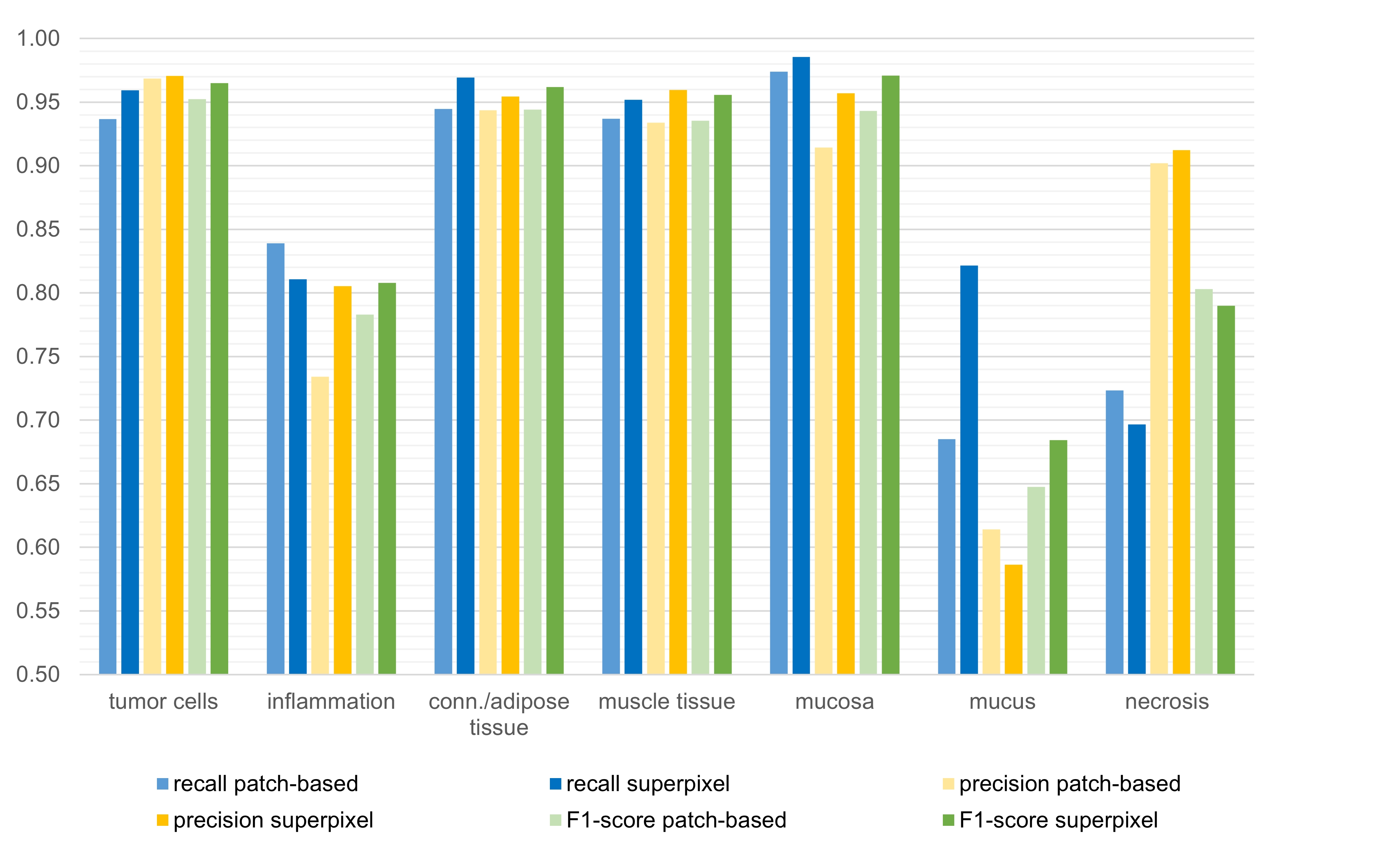}
    \caption{\label{fig:metrics}Comparison between patch-based and superpixel-based approach by class-wise recall, precision and F$_1$~score.}
\end{figure}

From the confusion matrices class-based recall and precision values are calculated, which are displayed in Figure~\ref{fig:metrics}. The superpixel-based approach yields an overall accuracy of 95.7~\% compared to 93.8~\% obtained with the patch-based approach. The improvement in accuracy has been tested for statistical significance using the two-matched-samples t-test based on the 29 slide-wise classification accuracies and has been verified on a confidence interval of 99~\%. Due to differences in the background detection which is performed per superpixel and respectively per patch, the sum of classified pixels slightly differs between the two approaches. Figure~\ref{fig:metrics} shows an improvement of the classification measures with the superpixel approach compared to the patch-based approach. The average improvement in recall is 0.022, 0.019 for precision and 0.018 for the F$_1$~score. Whilst this improvement can be observed for all classes with larger annotation areas, performance sometimes decreased for inflamed, necrotic and mucous areas. One possible reason for this might be that these classes constituted very fine annotations. The chosen superpixel size sometimes creates clusters too coarse to accurately represent these minute structures.

\begin{figure}[!ht]
    \centering
    \includegraphics[width=\columnwidth]{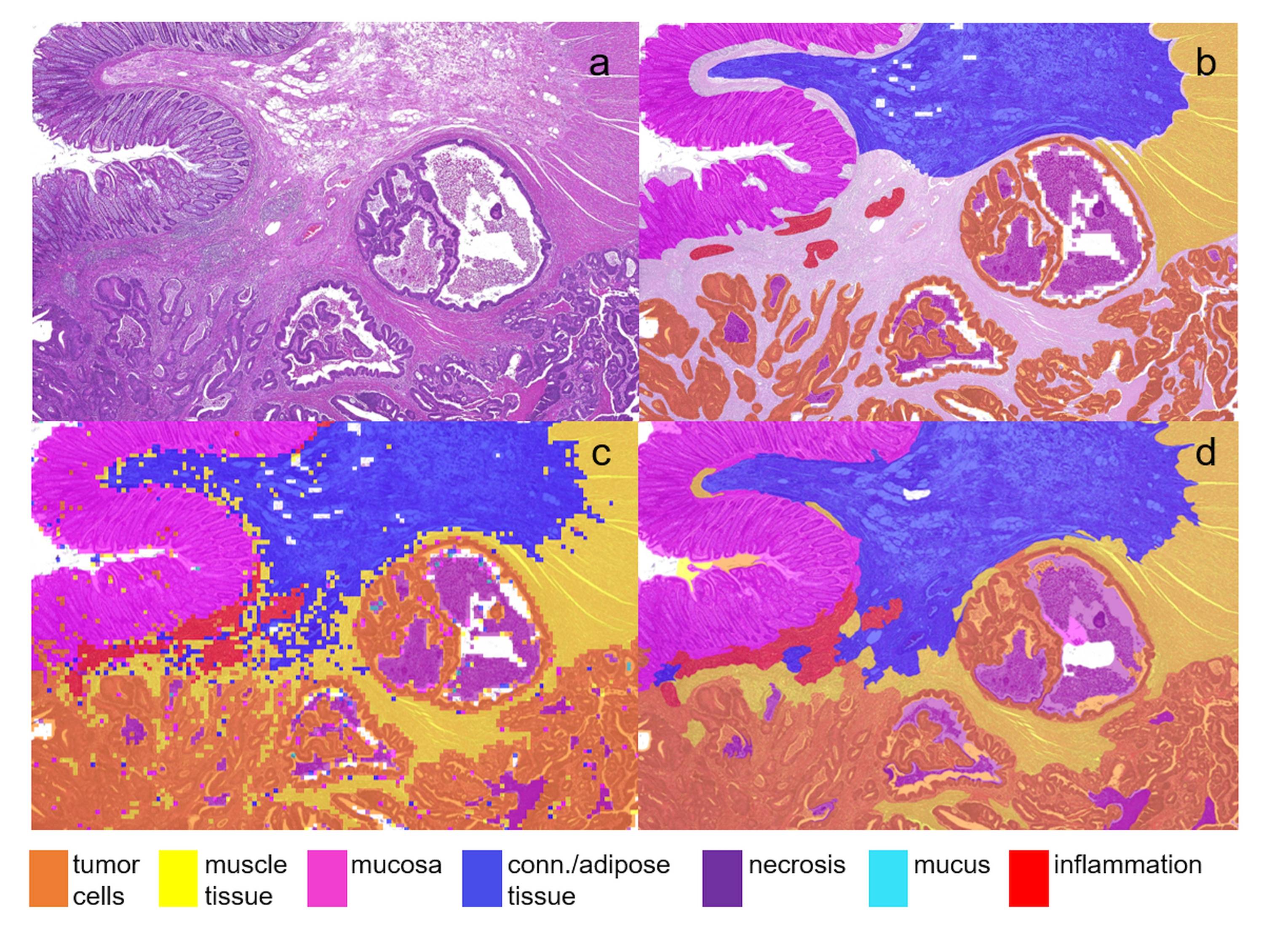}
    \caption{\label{fig:visual_results}Cartography results - a) Original section, b) Ground truth hand-annotation, c) Patch-based output, d) Superpixel-based output.}
\end{figure}

Figure~\ref{fig:visual_results} visualizes the cartography outputs of the compared approaches. Overall, the non-overlapping patch-based image analysis yields checkered classification outputs with many interruptions of connected components due to individual misclassifications. A prior segmentation into superpixels, on the other hand, yields smoother results which follow biological structures. It can be seen, that the larger tissue classes are detected accurately and also smaller structures, e.g. inflammations and necrotic areas, are classified correctly in most of the cases. However, this example also highlights limitations of the algorithm, where structures become too small to be accurately represented by the superpixels, e.g. small necrotic areas of comedo necrosis, which is in correspondence with the decrease in recall for necrosis compared to the patch-based approach. This drawback could be countered by choosing a smaller average superpixel size, albeit, only at the cost of higher computation times. The relatively large superpixel size also causes tumor cell classifications to be rather generous and incorporate surrounding tumor stroma. If a precise tumor/stroma separation is intended, the superpixel-based classification approach could be followed by a separate cell-detection-algorithm or simply a second refinement run of the superpixel segmentation and classification restricted to only the tumor areas.

Using an NVIDIA GeForce GTX 1060 \acs{gpu} and TensorFlow 2.2, the standard classification-based segmentation approach with non-overlapping patches resulted in computation times of 12.8 $\pm$~5.3 minutes per \ac{wsi}. The superpixel-based segmentation pipeline achieved classification times of 6.7~$\pm$~2.8 minutes with an additional 47~$\pm$~18 seconds for the \ac{slic} clustering resulting in an overall run-time of 7.5~$\pm$~3.0 minutes per \ac{wsi}. Thereby, an average acceleration of 41~\% could be achieved by the proposed image analysis approach. This acceleration is mainly the result of restricting the number of classified patches per superpixel. Without restriction the classification time increases to 13.4~$\pm$~5.5 minutes and the overall run-time including \ac{slic} clustering to 14.2~$\pm$~5.7 minutes. This is slower than the patch-based approach but yields the highest overall accuracy with 96.0~\% which is an improvement of 0.3 percentage points compared to the the superpixel cartography with restriction of the classified number of patches per superpixel. 

When comparing computation times, it has to be considered that the patch-based approach was performed in the fastest possible way by using non-overlapping patches. Standard patch-based approaches, however, use overlapping image patches and interpolate classification results. When choosing an overlap of half the patch dimension the number of overall classifications already increases from n x n to (2n-1) x (2n-1). Even when using fast scanning architectures for avoiding redundant computations in overlapping image regions, the overall computational costs are assumed to further increase when using overlaps. This underlines the benefit of the proposed clustering prior to classification even further. 

\subsection{Introduction of rejection class based on classification confidence}
Aiming to minimize the effect of misclassifications on the final cartography output, we attempt to detect superpixels with uncertain classification results. This way, a rejection label can be assigned to these superpixels. Our hypothesis is that the remaining classification results are more reliable and therefore yield a higher overall accuracy as well as average class-wise precision, recall and F$_1$~score. This is done at the expense that unclassified areas are introduced which are not included in the calculation of classification quality measures. Superpixels with a confidence lower than a defined threshold are assigned to the rejection class and hence all pixels (resolution: 3.54~$\mu$m x 3.54~$\mu$m / pixel) inside them as well. All pixels of the remaining superpixels are evaluated as before (see \ref{sec:evaluation_methods_cartography}). As a consequence of the rejection of unsure pixels the number of classified pixels and therefore the number of correct and false predicted pixels decreases.

We compared the confusion matrix and classification metrics without and with rejection of uncertain superpixels. As rejection threshold we have chosen 0.1, which means that all superpixels with a $C_{diff}^{votes}$  smaller than 0.1 are assigned to the rejection class. In total, 1.3~\% of the pixels were rejected. The number of total true predictions decreases by 0.8~\% compared to the classification without rejection while the number of false predictions decreases by 11.8~\%. Overall, 1.9 million pixels that were correctly classified are discarded due to a low confidence value and 1.3 million pixels that were incorrectly classified. The overall accuracy increases to 96.1~\% compared to 95.7~\% without rejection of superpixels. Likewise, there is an improvement for all classes in precision (average 0.009), recall (average 0.007) and F$_1$~score (average 0.009). The highest impact is obtained for classes that are usually distributed over the whole tissue sample and cover very small sections like necrosis, inflammation and mucus. These results support our hypothesis that the remaining classification results are more reliable at the expense of introducing areas without classification. Therefore, it depends on the application which aspect is prioritized.

\begin{figure}[!ht]
    \centering
    \includegraphics[width=\columnwidth]{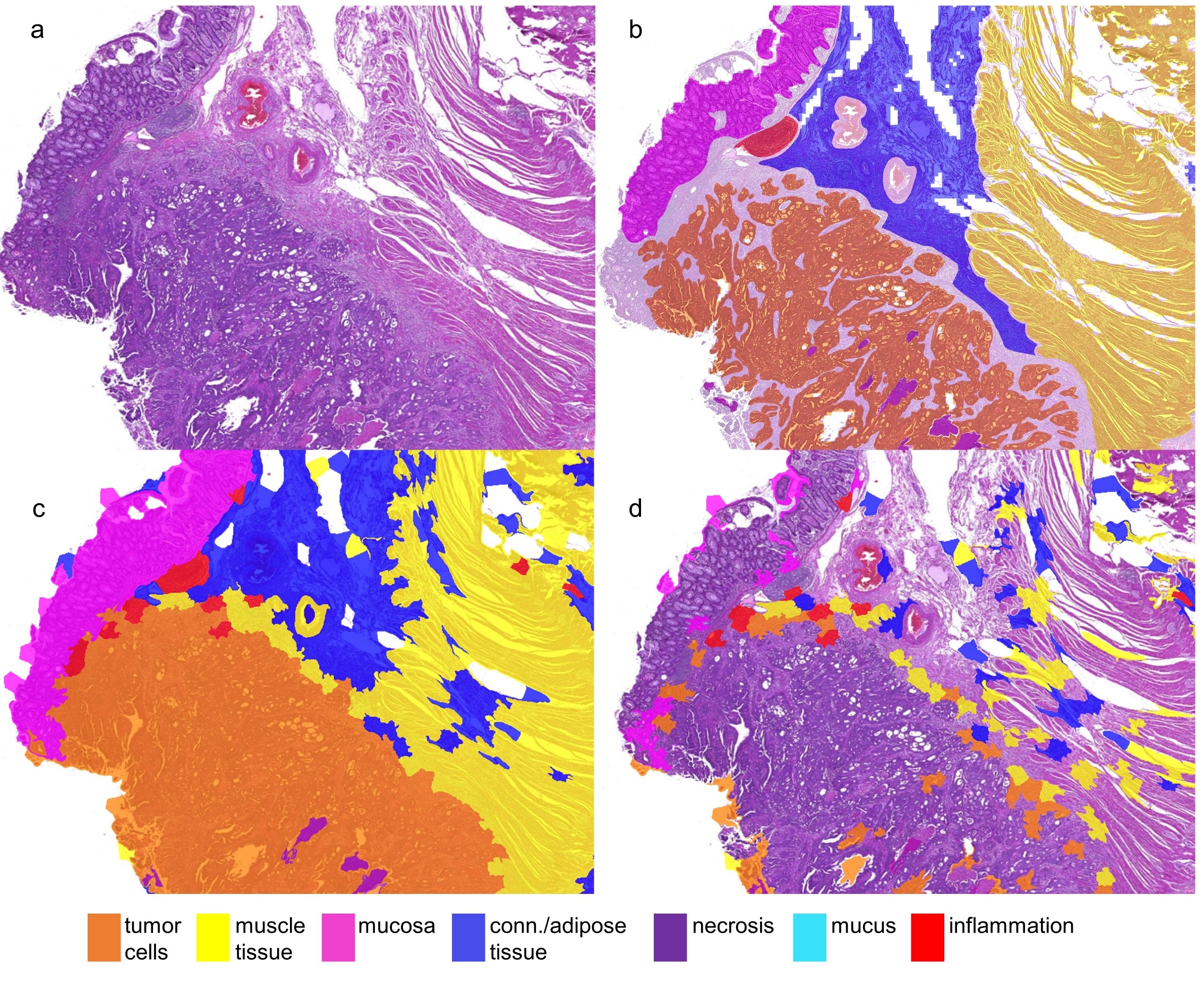}
    \caption{\label{fig:uncertain_SP}Example for uncertain superpixels based on $C_{diff}^{votes}$ with a threshold of 0.45 - a) Original section, b) Ground truth hand-annotation, c) Cartography results, d) Only superpixels with uncertain classification results are marked. Especially superpixels containing a high amount of background pixels or in the transition of two tissue types tend to show uncertain classifications.}
\end{figure}

Besides the quantitative evaluation, the question arises which areas in a \ac{wsi} tend to achieve uncertain classifications. We only touch upon this question with one qualitative example: In Figure~\ref{fig:uncertain_SP}d superpixels with uncertain classification results (based on $C_{diff}^{votes}$ with a threshold of 0.45) are highlighted. This example reveals two typical constellations that lead to an uncertain classification. Superpixels containing a high amount of background pixels, e.g. located at or nearby fissures or at the rim of the tissue section, tend to be misclassified. The same applies to superpixels in the transition of two tissue types, e.g. located near the invasive margin or slightly inflamed tissue. Moreover, ground truth annotations are only provided for regions that can be assigned clearly to one tissue type except for the tumor cell class. Here it was not feasible to annotate each small necrotic area which can be seen in Figure~\ref{fig:uncertain_SP}b.

\subsection{Tumor area}
\begin{figure}[!ht]
    \centering
    \includegraphics[width=\columnwidth]{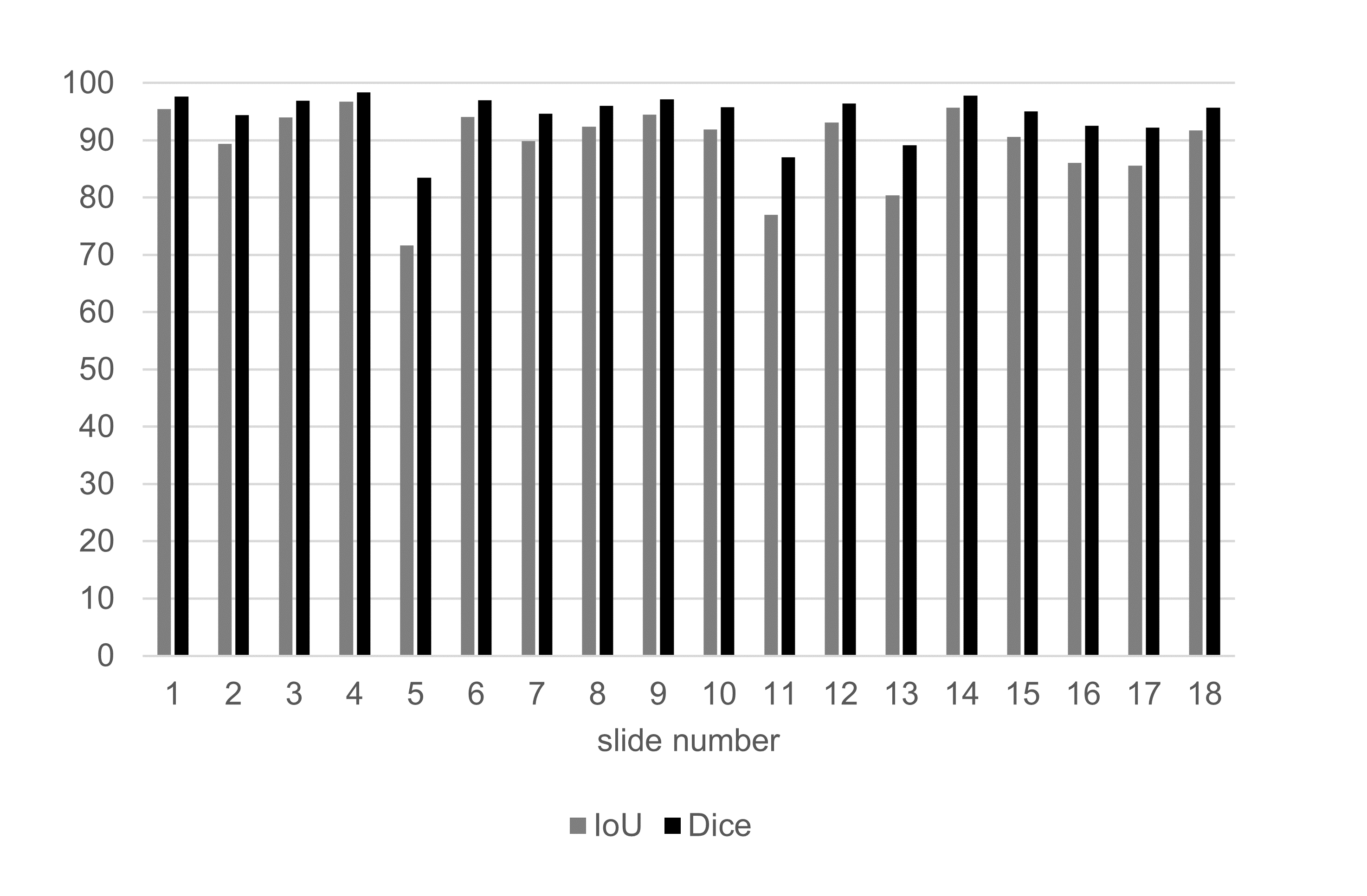}
    \caption{\label{fig:iou_dice_tumor}IoU and Dice measure of estimated and annotated tumor area for all 18 slides of dataset B.}
\end{figure}

Dataset B was used to evaluate the computation of the tumor area. On average, the estimated and the annotated tumor area differ by 6~\%  with a mean \ac{iou} of 89.4~\% and a mean Dice coefficient of 94.3~\% (per slide results in Figure~ \ref{fig:iou_dice_tumor}). 

\begin{figure}[!ht]
    \subfloat[Comparison of tumor area (from left to right: slide numbers 1,3,8,16 (see Figure~\ref{fig:iou_dice_tumor}). Misclassifications are largely located at the tumor boundary.]{ \includegraphics[width=\columnwidth]{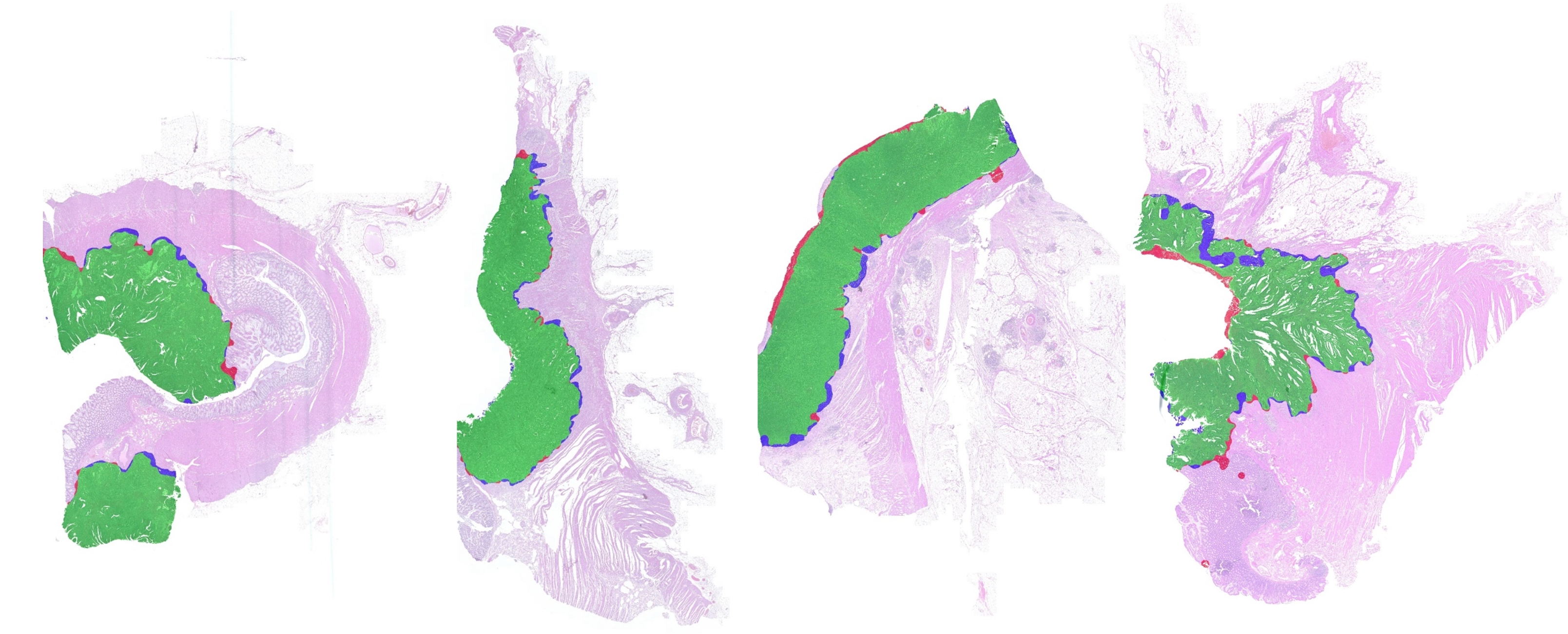}}
    \hfill
    \subfloat[Comparison of estimated and annotated tumor area showing the examples with the highest deviations (from left to right: slide numbers 5,11,13 (see Figure~\ref{fig:iou_dice_tumor}).]{ \includegraphics[width=\columnwidth]{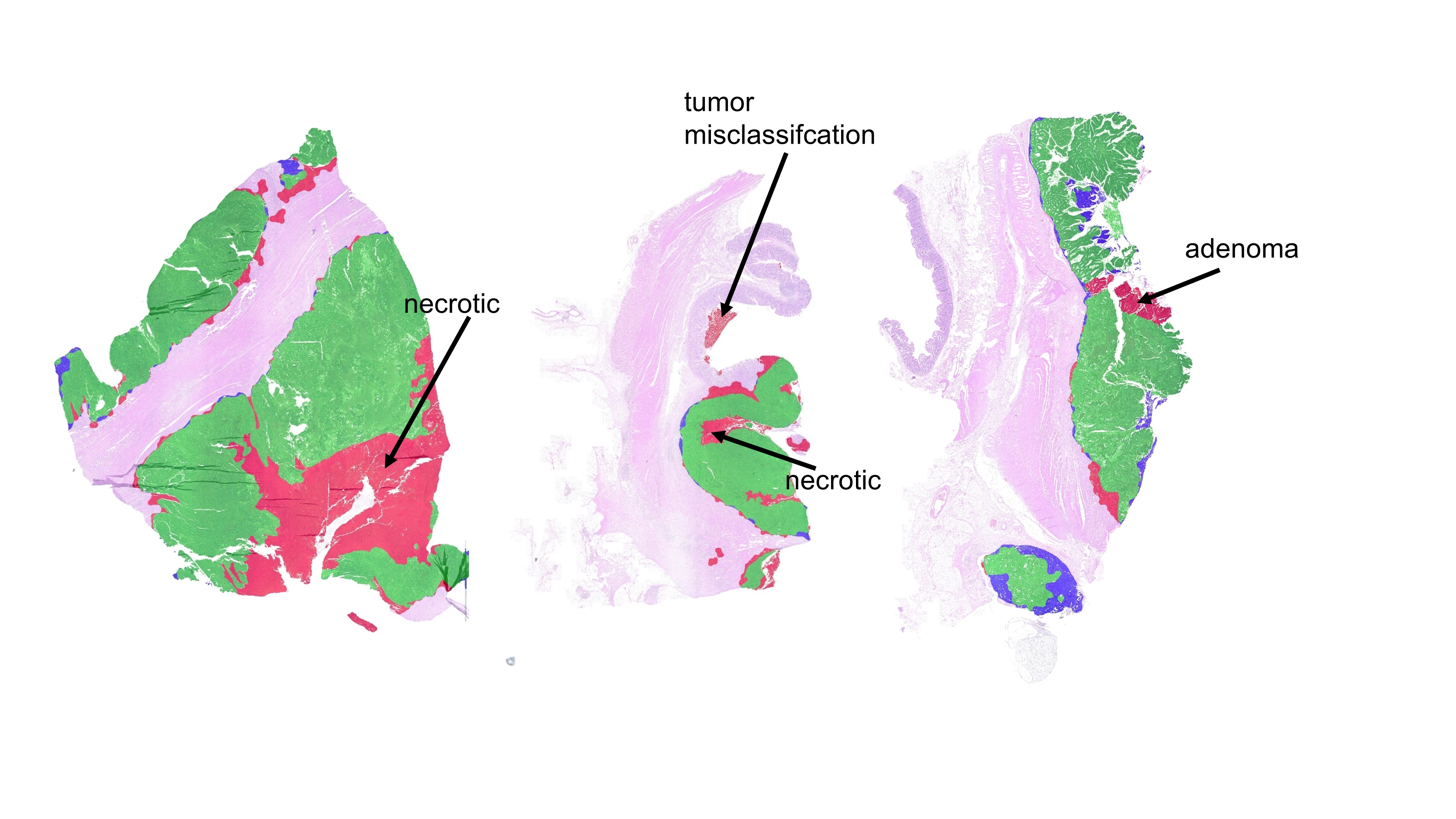}}
    \caption{\label{fig:tumor_area} Green: areas correctly identified as tumor (TPs); red: areas mistaken as tumor (FPs); blue: tumor tissue missed by the classifier (FNs).}
\end{figure}

Figure~\ref{fig:tumor_area}a depicts examples of evaluation results, where green overlays resemble tumor areas that have been found correctly (TPs), red marks areas that were mistaken as tumor (FPs) and blue indicates tumor annotations not detected by the algorithm (FNs). It can be seen, that most misclassifications are located at tumor boundaries. Especially necrotic areas adjacent to the lumen were included in the tumor area for our approach, but have been excluded by the pathologist. On the contrary, at the invasive margin our approach misses some tumor areas.

\begin{figure}[!ht]
    \centering
    \includegraphics[height=4cm]{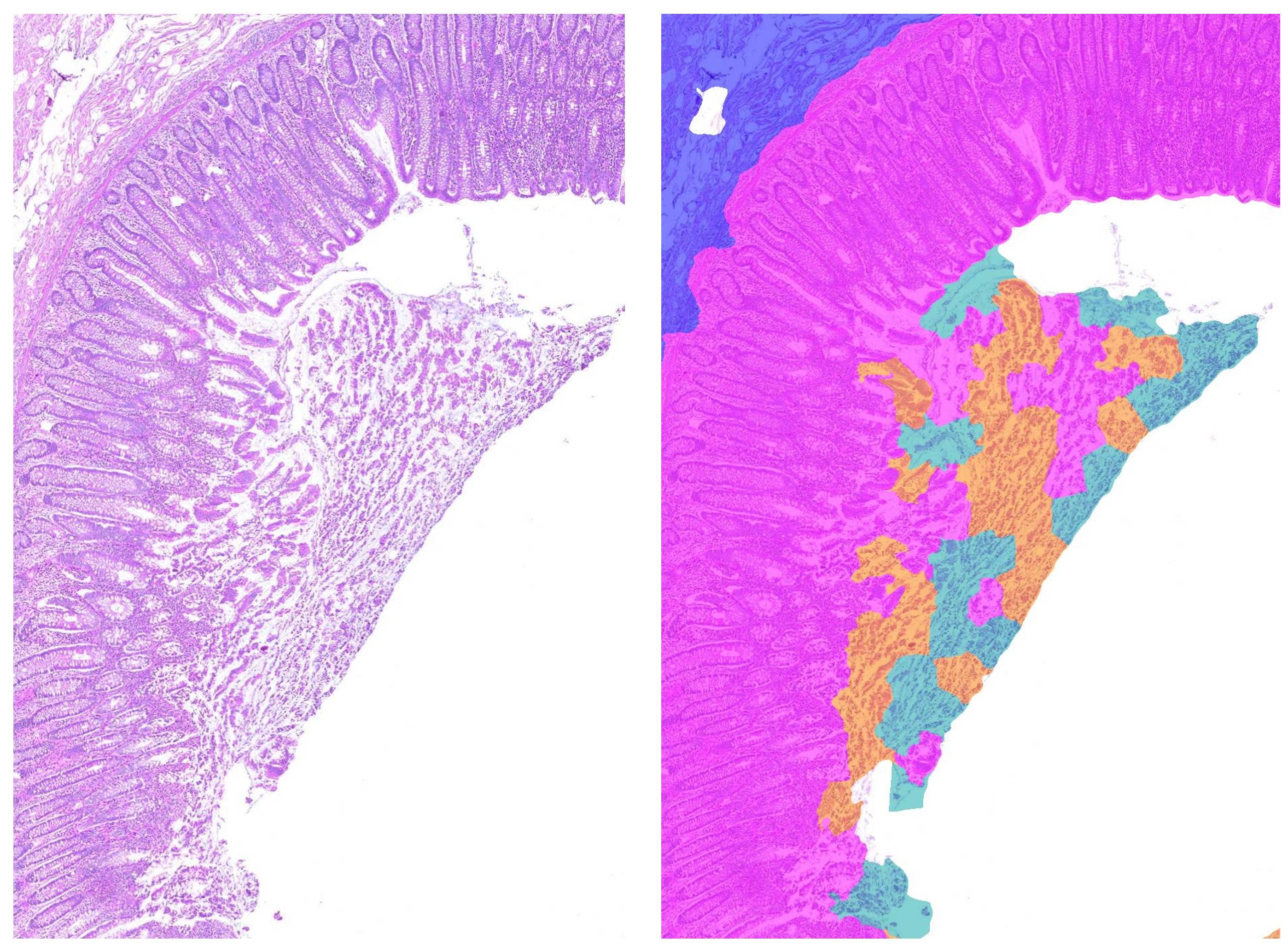}
    \caption{\label{fig:tumor_missclassification}The mixture of debris and destroyed mucosa tissue is classified as tumor (orange) and mucus (turquoise) and leads to a deviation in tumor area in slide number 11.}
\end{figure}

Looking at the slide results in detail, however, a few \acp{wsi} contain larger misclassified regions. Three examples are visualized in Figure~\ref{fig:tumor_area}b. One main source for deviations are again necrotic areas. In our approach all adjacent necrotic areas are incorporated into the tumor area. This technically defined rule cannot perfectly represent the pathologist's annotation (ground truth) in individual cases, as it cannot sufficiently reflect the biological and complex morphological nature of the tumor. Moreover, two sections contained adenomas that were classified as tumor. In rare cases tumor misclassifications occurred e.g. in areas containing debris and destroyed mucosa tissue (see Figure~\ref{fig:tumor_missclassification}).

\subsection{Invasive margin}
\begin{figure}[!ht]
    \centering
    \includegraphics[width=\columnwidth]{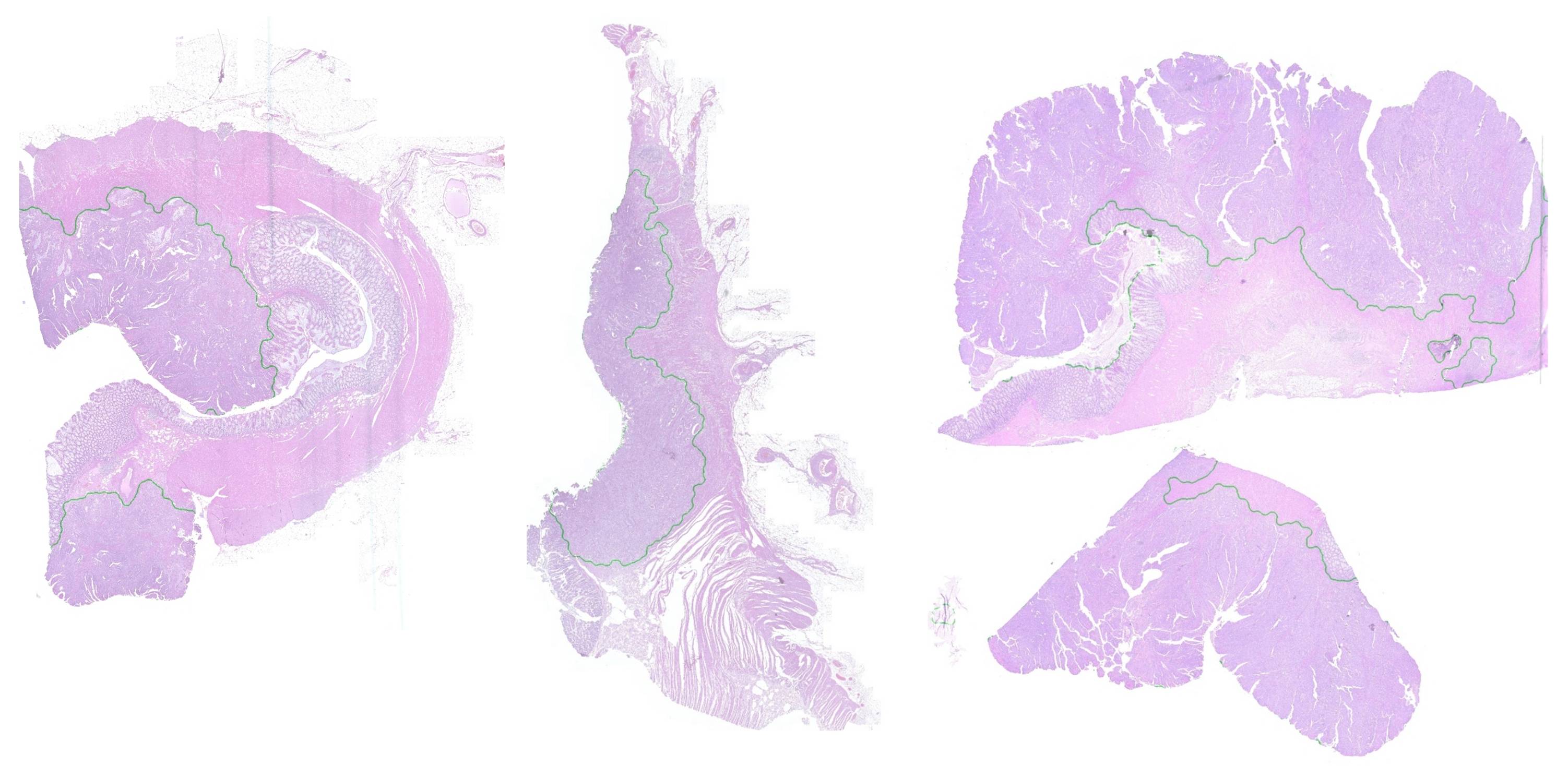}
    \caption{\label{fig:invasive_margin}Examples of automatically generated invasive margins (marked in green).}
\end{figure}

By growing the tumor area evenly by a defined distance towards the surrounding healthy classes, the tumor invasive margin can automatically be generated (see Figure~\ref{fig:invasive_margin}). The generated margins of all slides of dataset B are qualitatively evaluated by two pathologist using a point-based grading system from 1 to 5 (1 $\hat{=}$ very good, 5 $\hat{=}$ insufficient). On average, the margins were rated 1.6 composed of 18 ratings as “very good”, 15 ratings as “good” and three as “satisfying”. The two pathologists were in correspondence for 13 \acp{wsi} and their judgments only differed by one point for five \acp{wsi}. These first qualitative results seem promising and could enable further analysis, e.g. the determination of the invasion depth or quantifying inflammation within the invasive margin.

\subsection{Tumor composition}
\begin{figure}[!ht]
    \centering
    \includegraphics[width=\columnwidth]{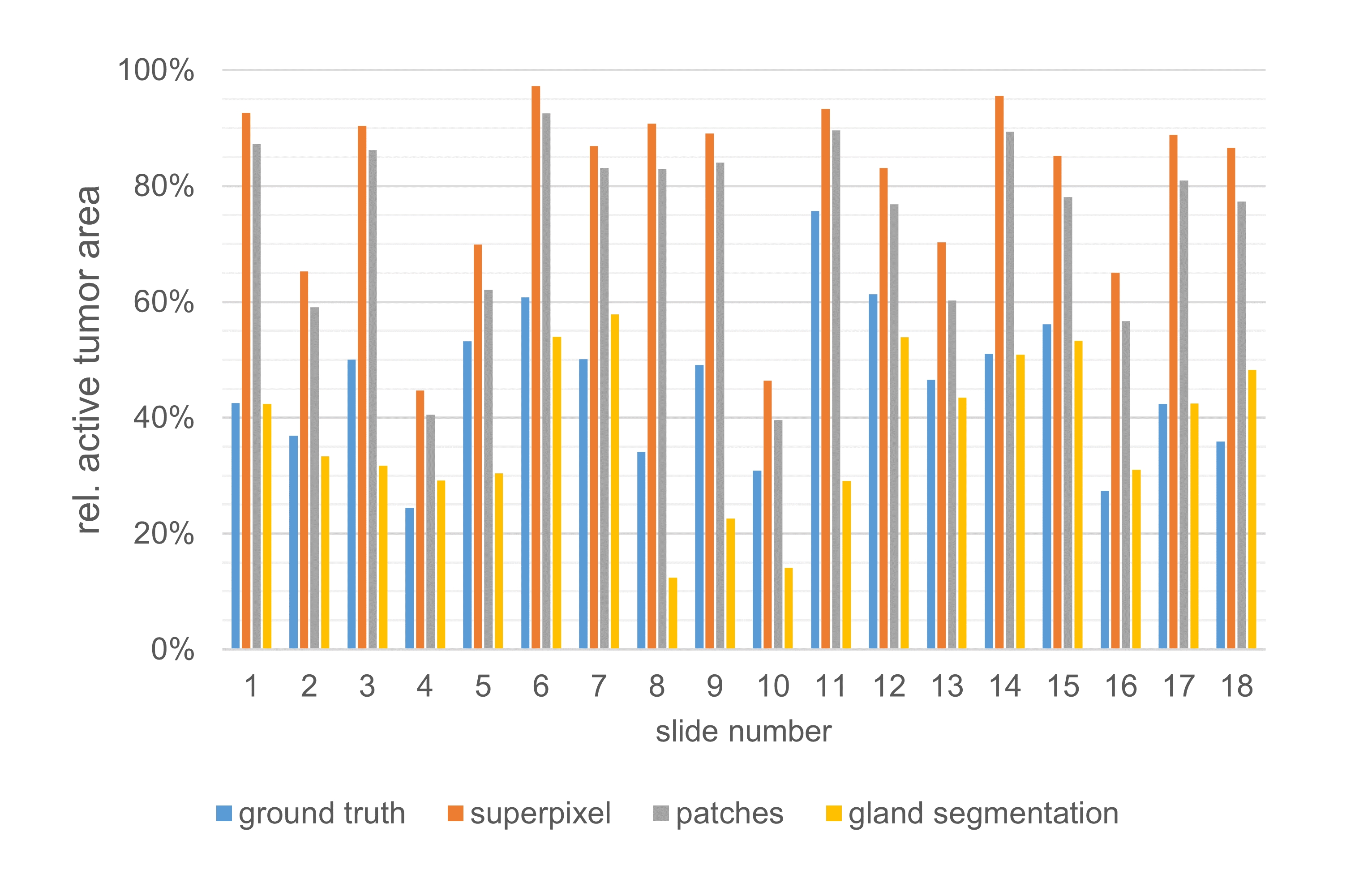}
    \caption{\label{fig:active_tumor_area}Active tumor area relative to annotated ground truth tumor area calculated with different approaches.}
\end{figure}

Using dataset B, the tumor composition is evaluated by computing ratios of tumor cells (Figure~\ref{fig:active_tumor_area}), necrosis and mucus within the ground truth tumor area. The results in Figure~\ref{fig:active_tumor_area} show that both the superpixel approach and the patch-based approach overestimate the active tumor area for every slide. However, the average deviation is smaller in the latter case. The best estimation is obtained with the gland segmentation approach.

In order to analyze these results further, the slides of dataset B have been divided into subsets according to their tumor grading. Table~\ref{tab:active_tumor_area} breaks down the deviation of estimated active tumor area from the ground truth for each subset as mean over the subset. For tumors with grade 1 and grade 2 the gland segmentation approach provides good estimations of the active tumor area. As expected, the accuracy decreases with tumor belonging to grade 3 where the growth becomes diffuse and gland structure is destroyed.

\begin{table}[!ht]
    \caption{\label{tab:active_tumor_area}Comparison of different methods for the determination of active tumor area. The average deviation between the calculated relative active tumor area and the ground truth relative active tumor area is given. Slides are assigned to the set “grade 3” as soon as parts with grade 3 are present.}
    \centering
    \resizebox{\columnwidth}{!}{%
    \begin{tabular}{|c|c|c|c|}
        \hline
         grade & superpixel & patch-based & gland \\
         &&&segmentation \\
         \hline
         1 - 2 & 34.7~\% & 28.3~\% & 3.7~\%\\
         3 & 33.2~\% & 26.9~\% & 21.1~\%\\
         \hline
         Dataset B & 34.0~\% & 27.7~\% & 11.4~\%\\
         \hline
    \end{tabular}}
\end{table}

\begin{figure}[!ht]
    \centering
    \includegraphics[width=\columnwidth]{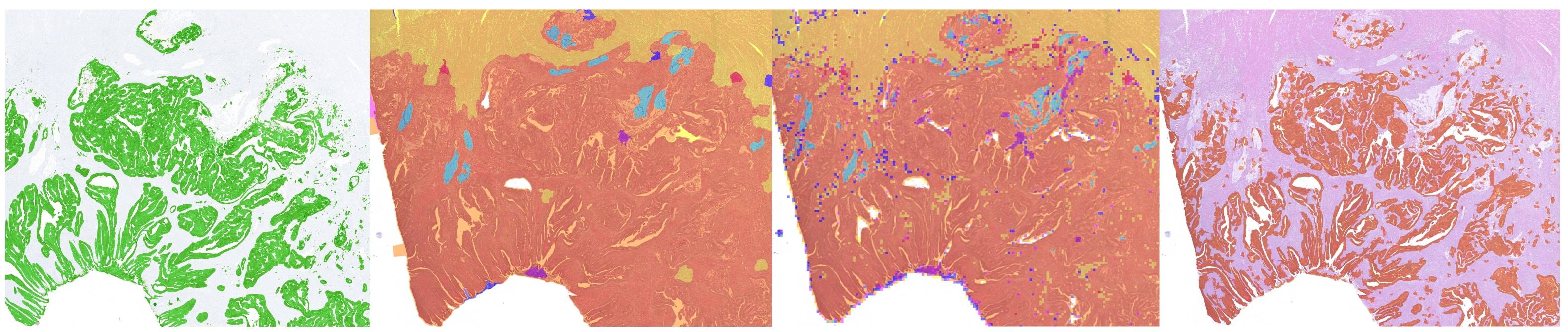}
    \caption{\label{fig:active_tumor_area_example1}Comparison of active tumor areas obtained with different approaches for slide number 1 with a tumor of grade 2. From left to right: ground truth segmentation (green) of active tumor area in IHC staining. Cartography results in \ac{he} staining by superpixel approach and patch-based approach. Gland segmentation results. Active tumor area is depicted in orange for all three approaches.}
\end{figure}

\begin{figure}[!ht]
    \centering
    \includegraphics[width=\columnwidth]{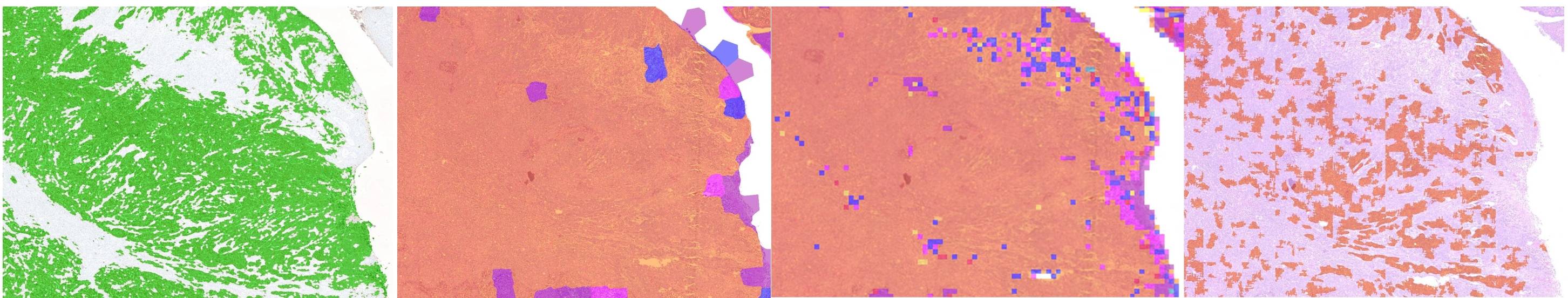}
    \caption{\label{fig:active_tumor_area_example2}Comparison of active tumor areas obtained with different approaches for slide number 11 with a tumor of grade 3. From left to right: ground truth segmentation (green) of active tumor area in IHC staining. Cartography results in \ac{he} staining by superpixel approach and patch-based approach. Gland segmentation results. Active tumor area is depicted in orange for all three approaches.}
\end{figure}

Figure~\ref{fig:active_tumor_area_example1} shows the detected active tumor area (marked in orange) for a well differentiated tumor (grade 2, slide number 1) for all three approaches. This example illustrates that the superpixel approach overestimates the active tumor area due to misclassification of tumor stroma as tumor cells. The patch-based approach shows a similar behavior, however, with smaller deviations to the ground truth. The gland detection approach is in good correspondence with the ground truth segmentation. The limitation of this approach is evident in the second example (Figure~\ref{fig:active_tumor_area_example2}) showing a tumor with grade 3 (slide number 11). In this example the estimated active tumor area deviates significantly from the ground truth area (Table~\ref{tab:active_tumor_area}). On the contrary, the deviation to the ground truth for the superpixel and patch-based approaches seems to be independent of the grade of the tumor.

Besides the active tumor area, the ratio of necrosis and mucus area within the tumor area are additional relevant parameters for characterization of the tumor micro-environment. Both, the patch-based and superpixel approach show similar results here with a slight superiority of the patch-based approach for the determination of the necrotic area (see Table~\ref{tab:necrosis_mucus}). Because the average superpixel size (0.048~mm$^2$ on dataset B) is significantly bigger than the patch size (0.002~mm$^2$) necrotic areas, which are oftentimes only small islands between tumor cells, seem to be better captured by the patches than the superpixels. 

\begin{table}[!ht]
    \caption{\label{tab:necrosis_mucus}Comparison of superpixel and patch-based approach for the determination of necrotic and mucus area. Necrotic areas were present in all of the slides. However only five slides contained mucus areas. Therefore, the average deviation for the relative mucus area is calculated once for all slides and once only for slides containing mucus.}
    \centering
    \resizebox{\columnwidth}{!}{%
    \begin{tabular}{|l|c|c|}
        \hline
        average deviation & superpixel & patch-based \\
        \hline
        rel. necrotic area & 1.77~\% & 1.22~\%\\
        rel. mucus area (all WSIs) & 0.19~\% & 0.24~\%\\
        rel. mucus area (only WSIs & 0.56~\% & 0.55~\%\\
        containing mucus) & & \\
        \hline
    \end{tabular}}
\end{table}

\section{Conclusion}
In this work we presented an approach for histology whole-slide cartography using superpixels by the example of colon carcinomas. Our work was motivated by a feasibility of the developed method in a clinical setting. Even though, regarding granularity of segmentation outputs, encoder-decoder-based approaches are sometimes considered superior to patch-based approaches, they oftentimes require powerful hardware not attainable in a clinical setting. Therefore, our work focused on increasing the efficiency of a patch-based cartography, which can easily be transferred to e.g. a pathology institute and ensures fast inference. This increased efficiency could be obtained by pre-segmenting the input image into superpixels and only classifying a subset of patches within these superpixels.

The evaluation results on our test set composed of 29 \acp{wsi} show a superiority of our approach compared to a classical patch-based approach for overall accuracy with an increase from 93.8~\% to 95.7~\% as well as computing-time with an average speed-up of 41~\% resulting in an average overall run-time per \ac{wsi} of 7.5 minutes. The speed-up is mainly achieved by limiting the number of classified patches within each superpixel. This patch restriction only results in a marginal decrease in accuracy of 0.3 percentage points compared to the unrestricted approach. These results indicate that the superpixel clustering already segments the \ac{wsi} into regions belonging to the same tissue type. Only when this requirement is fulfilled can accurate cartography results be obtained. The limitation of our approach lies in the relatively large size of superpixels compared to patches. On our test set one superpixel on average covers 0.05~mm$^2$. Compared to fine-grained structures, such as small necrotic areas within the tumor, this size is too big to correctly capture these areas. This is also reflected in e.g. a lower recall for necrosis. Another limitation lies in the manual annotations. Accurate and complete annotation of these fine-grained structures are also a challenge for the human annotator. Therefore, wherever possible, an alternative generation of the ground truth should be preferred, e.g. based on segmentation in immunohistochemically stained sections. Moreover, one has to keep in mind that there is a general problem with the quantitative assessment of cartography results. Although using seven tissue classes, there are still areas that cannot be clearly assigned to one of these classes. These non-annotated areas are not included in the quantitative evaluation. Therefore, from our point of view, it is important to always apply the developed approaches to complete \acp{wsi} and check the cartography results in these areas at least qualitatively.

The key difference of our method compared to other superpixel-based approaches for histopathology images is the one-to-many relationship between superpixels and corresponding image patches. In our setup, a superpixel contains on average twenty image patches of which we classify a random subset. Utilizing the fact that a superpixel class-label is inferred from a set of multiple individually classified patches, we investigated a measure for quantifying the uncertainty of a superpixel classification derived from the votes of the patches within the superpixel. This measure was suited to decrease the relative number of incorrect predictions at the cost of introducing unclassified tissue areas and a rejection of correct predictions, but to a smaller extent. Moreover, applying our introduced uncertainty measures to \acp{wsi} and visualizing uncertain superpixels enables a plausibility check of the approach. As expected, classification results of superpixels in the transition of two tissue types, e.g. located near the invasive margin, tend to be unsure. The uncertainty measurement also facilitates an automatic improvement by e.g. partitioning them into smaller segments and re-classifying these superpixels or applying other pixel-wise segmentation methods within these areas.

Whole-slide cartography by itself offers only limited support to the pathologist but provides a basis for subsequent analysis operations that can predict various medical endpoints. Within this work we used the cartography results to determine the tumor area and composition as well as to derive the invasive margin. While the tumor area is in good agreement with the ground truth, the tumor composition analysis highlights weaknesses of the approach. Again, due to the size of our superpixels the separation between finely grained active tumor cells and tumor stroma is not adequate. However, a combination with further methods (in our case gland segmentation for well differentiated tumors) yields good results with an average deviation of only 11.4~\%.

Being able to reliably detect and outline the tumor area is very valuable from a clinical perspective. In a routine workflow, such a functionality could be used as an assistance system that draws the pathologist’s attention to a specific region. Alternatively, such a system could be introduced as a quality control mechanism that provides a second opinion. Another potential application is the insurance that samples for molecular testing are taken from an area that actually contains a high ratio of tumor cells. In the context of Computational Pathology, it has been shown in recent literature that it is possible to predict genetic alterations directly from \acp{wsi} \citep{kather2019,kather2020}. A pre-requirement here is that only tumor-areas are analyzed. In the context of colon carcinoma, an example would be the detection of microsatellite instability (MSI) to validate the presence of Lynch syndrome. The tumor composition in terms of the ratios of active tumor cells, tumor stroma, necrosis and mucus has been shown to be of prognostic relevance \citep{huijbers2013} and could at least for well differentiated tumors be assisted by the proposed approach.

\section*{Acknowledgments}
This work was supported by the Bavarian Ministry of Economic Affairs, Regional Development and Energy through the Center for Analytics – Data – Applications (ADA-Center) within the framework of ``BAYERN DIGITAL II'' (20-3410-2-9-8). This work was partially supported by the Federal Ministry of Education and Research under the project reference numbers 16FMD01K, 16FMD02 and 16FMD03. Parts of the research has been funded by the German Federal Ministry of Education and Research (BMBF) under the project TraMeExCo (011S18056A).

We also want to thank Christa Winkelmann and Natascha Leicht regarding technical issues in the laboratory and regarding sectioning and staining. We want to thank Nicole Fuhrich and the working group for digital pathology at the institute of pathology like Tatjana Kulok and Jonas Plum for assistance regarding all scans of the cohort.


\nocite{*}
\bibliographystyle{unsrtnat85}
\bibliography{refs}

\begin{acronym}
\acro{wsi}[WSI]{whole-slide-image}
\acro{he}[H\&E]{Hematoxylin \& Eosin}
\acro{cnn}[CNN]{Convolutional Neural Network}
\acro{fcn}[FCN]{Fully Convolutional Neural Network}
\acro{gpu}[GPU]{graphics processing unit}
\acro{iou}[IoU]{Intersection over Union}
\acro{svm}[SVM]{Support Vector Machine}
\acro{uker}[UKER]{University Hospital Erlangen}
\acro{ihc}[IHC]{immune histochemical staining}
\acro{slic}[SLIC]{Simple Linear Iterative Clustering}
\acro{til}[TIL]{tumor-infiltrating lymphocyte} 
\end{acronym}

\end{document}